\documentclass{article}
\usepackage[title]{appendix}
\usepackage{arxiv}

\usepackage[utf8]{inputenc} 
\usepackage[T1]{fontenc}    
\usepackage{hyperref}       
\usepackage{url}            
\usepackage{booktabs}       
\usepackage{amsfonts}       
\usepackage{nicefrac}       
\usepackage{microtype}      
\usepackage{lipsum}		
\usepackage{graphicx}
\usepackage{natbib}
\usepackage{doi}
\usepackage{amsmath}
\usepackage{wrapfig}
\usepackage{appendix}

\title{Structure to Property: Chemical Element Embeddings and a Deep Learning Approach for Accurate Prediction of Chemical Properties}

\author{
    \href{https://orcid.org/0000-0002-3429-7493}{\includegraphics[scale=0.06]{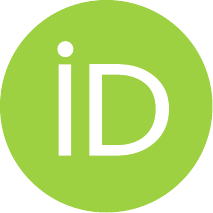}\hspace{1mm} Shokirbek Shermukhamedov}\textsuperscript{1,*}, 
    Dilorom Mamurjonova\textsuperscript{2}, 
    Michael Probst\textsuperscript{1,3} \\
    \\
    \textsuperscript{1}Institute of Ion Physics and Applied Physics, University of Innsbruck, 6020 Innsbruck, Austria \\
    \textsuperscript{2}Tashkent Chemical Technological Institute, 100011 Tashkent, Uzbekistan \\
    \textsuperscript{3}School of Molecular Science and Engineering, Vidyasirimedhi Institute of Science and Technology, \\21201 Rayong, Thailand \\
    \\
    *\textit{Author to whom any correspondence should be addressed.} \\
    E-mail: \href{mailto:shokirbek.shermukhamedov@uibk.ac.at}{shokirbek.shermukhamedov@uibk.ac.at}
}

\renewcommand{\shorttitle}{elEmBERT}

\begin{document}
\maketitle
\begin{abstract}
We introduce the elEmBERT model for chemical classification tasks. It is based on deep learning techniques, such as a multilayer encoder architecture. We demonstrate the opportunities offered by our approach on sets of organic, inorganic and crystalline compounds. In particular, we developed and tested the model using the \textit{Matbench} and \textit{MoleculeNet} benchmarks, which include crystal properties and drug design-related benchmarks. We also conduct an analysis of vector representations of chemical compounds, shedding light on the underlying patterns in structural data. Our model exhibits exceptional predictive capabilities and proves universally applicable to molecular and material datasets. For instance, on the Tox21 dataset, we achieved an average precision of 96\%, surpassing the previously best result by 10\%. 

\end{abstract}


\section{Introduction}

Due to their effectiveness in fitting experimental data and predicting material properties, machine learning models have found extensive applications in research on batteries \cite{Ng2020Predicting,Liu2020Machine}, supercapacitors \cite{Sawant2023Machine}, thermoelectric \cite{Iwasaki2019Machine-learning}, and photoelectric \cite{Akhter2019Review} devices, catalysts \cite{Toyao2020Machine}, and in drug design \cite{Vamathevan2019Applications}. In a 'second wave', deep learning models (DLMs) have exhibited remarkable potential in advancing the field of chemical applications. Word2vec \cite{Mikolov2013Efficient} DLMs have been used for processing textual chemical data extracted from academic articles. By representing chemical formulas as embeddings or vectors, non-obvious connections between compounds and chemical properties can be discovered. For instance, the mat2vec \cite{Tshitoyan2019Unsupervised} NLP model was able to predict materials with good thermoelectric properties even when these materials and their properties were not explicitly named in the original papers. Other NLP-inspired models such as Bag of Bonds \cite{Hansen2015Machine}, mol2vec \cite{Jaeger2018Mol2vec}, smiles2vec \cite{Goh2017SMILES2Vec}, SPvec \cite{Zhang2020SPVec} have used unsupervised machine learning and have been applied to chemical compound classification tasks, achieving remarkable results. These models hold immense potential for accelerating the discovery and design of materials with tailored properties.

In this regard, the type of input data is crucial for ML models. In chemistry, this could be chemical text data like in mat2vec or structural data. Chemical texts make it possible to use reference information about a compound \cite{Stanev2018Machine} such as molecular weight, melting point, crystallization temperature, and element composition. These types of inputs can in turn be used by general deep learning models with ELMO, BERT, and GPT-4 being the most famous examples. However, such models cannot directly capture 3D information from structural files.

One of the most common types of input data used for ML-based approaches is structural representation, which provides valuable information about the atomic environment of a given material. However, text-based data does not normally capture important structural features such as interatomic distances. Structural information is crucial for predicting material properties, as it is key to all pertinent physical and chemical characteristics. This can be understood in the same sense as the Born-Oppenheimer approximation, which in short states that atomic coordinates (and from them the potential energy) are all that is needed in chemistry. The challenge of linking structural information to material properties is commonly referred to as the "structure to property" task. Overcoming this challenge has the potential to greatly enhance our ability to predict and design novel materials with desired properties.

Structure can be translated into property by graph neural networks (GNN) or high-dimensional neural networks (HDNN) formalisms. GNNs transform graphs of molecules (or compounds) into node and edge embeddings, which can then be used for state-of-the-art tasks \cite{Chen2019Graph, Kong2022Density, Gori2005new, Zhang2020Molecular,Shui2020Heterogeneous, Fung2021Benchmarking, Xie2018Crystal, Guo2021Few-shot}. GNNs are efficiently applied for both chemical classification and regression tasks. However, as the size and complexity of molecular graphs increase, the computational requirements for GNNs also grow. Handling large graphs with many atoms or intricate structures can pose scalability challenges both in terms of memory usage and computer time. HDNNs based on converting Cartesian coordinates of atoms to continuous representations utilize techniques such as smooth overlap of atomic positions (SOAP) \cite{Bartók2013On}, many-body tensor representation (MBTR) \cite{Huo2022Unified}, or atomic centered symmetry functions (ACSF) \cite{Behler2021Four} to achieve the same goal. Message passing neural networks (MPNN) are a subgroup of HDNN that employ atomic positions and nuclear charges as input. The PhysNet \cite{Unke2019PhysNet} model serves as a typical example. In this model, atomic embedding encodes the atomic identifier into vector arrays, which are first initialized randomly and optimized during training. 

HDNNs demonstrate excellent performance on regression tasks. However, in classification tasks, GNNs dominate. Moreover, a disadvantage of HDNNs is their susceptibility to overfitting and their computational complexity. As the dimensionality increases, the possibility of overfitting rises due to the larger number of parameters, and the training time also increases.

Despite the increasing use of deep learning in computational chemistry, many aspects of NLP models have yet to be fully explored. One of them is the attention mechanism \cite{Vaswani2017Attention}, which allows the model to focus on specific parts of the input data when making predictions. It works by assigning different levels of importance or attention to different elements in the input sequence. The attention mechanism has been previously used in graph neural networks \cite{Louis2020Graph}. Additionally, the transformer approach, commonly known for its successful application in chemical GNNs, has not been fully extensively applied to HDNNs. The transformer consists of two distinct components: an encoder responsible for processing the input data and a decoder responsible for generating task-related predictions. In this paper, we introduce a new deep learning model for chemical compounds that utilizes the encoder part of the Transformer architecture. Specifically, our model incorporates local attention layers to capture properties of local atomic environments and then utilizes a global attention layer to make weighted aggregations of these atomic environment vectors to create a global representation of the entire chemical structure. From its components, we call this model 'elEmBERT' (\textbf{el}ement \textbf{Em}beddings and \textbf{B}idirectional \textbf{E}ncoder \textbf{R}epresentations from \textbf{T}ransformers).

In summary, the main aspects of our work are:
\begin{itemize}
    \item We use a transformer mechanism for binary and multilabel classification based on structural information.
    \item Our model is flexible, fast, and can be easily adapted to different types of datasets.
    \item Benchmarks show the state-of-the-art performance of our model for a variety of material property prediction problems, both involving organic and inorganic compounds.
\end{itemize}

\section{Methods}

\begin{figure*}[t]
\centering
\includegraphics[width=0.9\linewidth]{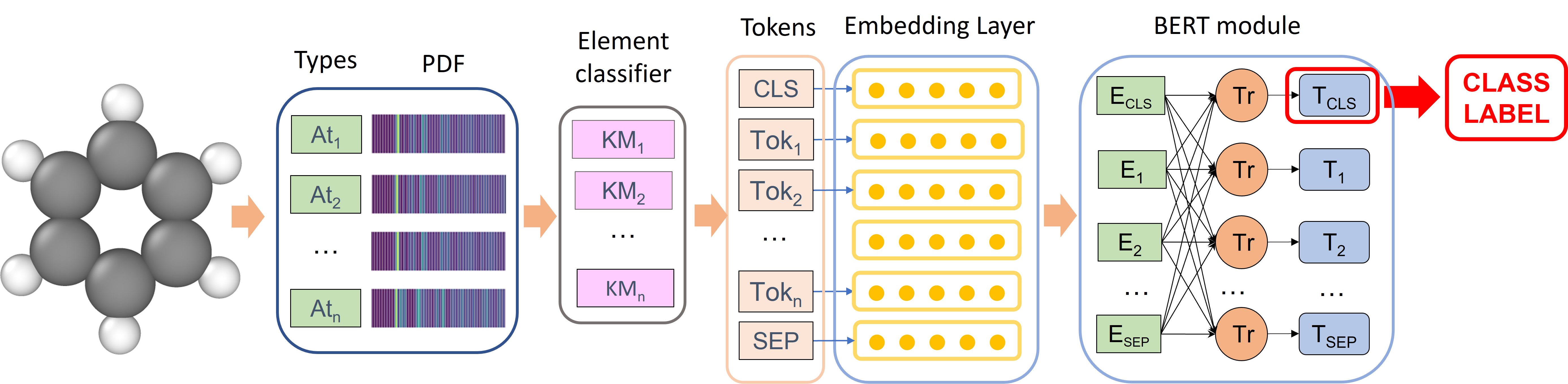}
\caption{\textbf{elEmBERT model architecture}. The initial step involves computing the pair distribution function for each element based on atom positions within the chemical compound. This information is then passed through the classifier model. Subsequently, the resulting sub-elements are converted into tokens, with additional tokens incorporated before input into the BERT module. The [CLS] token output vector from BERT is used for the classification task. }
\label{fig1} 
\end{figure*}
As input to the neural network (NN), we use atomic pair distribution functions (PDFs) and the atom types that compose the compounds. The PDF represents the probability of finding an atom inside a sphere with a radius $r_\text{cut}$ centered at a selected atom \cite{Billinge2019rise}. To prepare the training data, we calculate PDFs employing the ASE library\cite{Larsen2017atomic} with a cutoff radius of 10 Å \cite{Shermukhamedov2023Structure}. The second input for the NN consists of element embedding vectors. To achieve this, all elements in all compounds are mapped to integers (typically using the nuclear charge), creating an elemental vocabulary of size ${V_{size}=101}$. These embeddings are then passed to the BERT module.
BERT is a deep learning architecture originally designed for natural language processing (NLP) tasks. It employs a bidirectional transformer encoder to capture word context in sentences, allowing it to generate accurate text representations. BERT employs masked language modeling (MLM), where some tokens in a sentence are masked or replaced with a [MASK] token, and the model is trained to predict the original word based on the surrounding context. 

\subsection{Model architecture }
Our model architecture is illustrated in Fig. \ref{fig1}. It can use various combinations of embedding sizes, encoder layers, and attention heads. In chemical applications, the atomic composition of a compound can be equated to a sentence, with individual atoms serving as constituent tokens. Leveraging this analogy, we introduce four new tokens to the vocabulary: [MASK] for MLM, [UNK] for unseen tokens, [CLS] for classification, and [SEP] for separating two compounds. 
In standard BERT models positional embeddings play an important role by encoding the order of tokens in a sequence, which allows the model to capture sequential relationships between tokens. Since the Transformer approach is inherently permutation-invariant – treating different permutations of the same input sequence identically – there is no explicit encoding of token order. As well, in chemical compounds, the order of atoms does not affect the properties of the entire compound. Correspondingly, in our implementation, positional embeddings are intentionally omitted to preserve permutation invariance.
In chemical compounds, elements may exhibit varying oxidation states or formal charges, indicating the relative electron loss or gain during reactions. The interactions between these elements are non-uniform and follow specific patterns based on their neighboring atoms. For inorganic substances, these interactions typically appear as ionic interactions, signified by the oxidation state, while in organic substances, covalent bonding is prevalent.

\begin{figure}[h]
\centering
\includegraphics[width=0.8\linewidth]{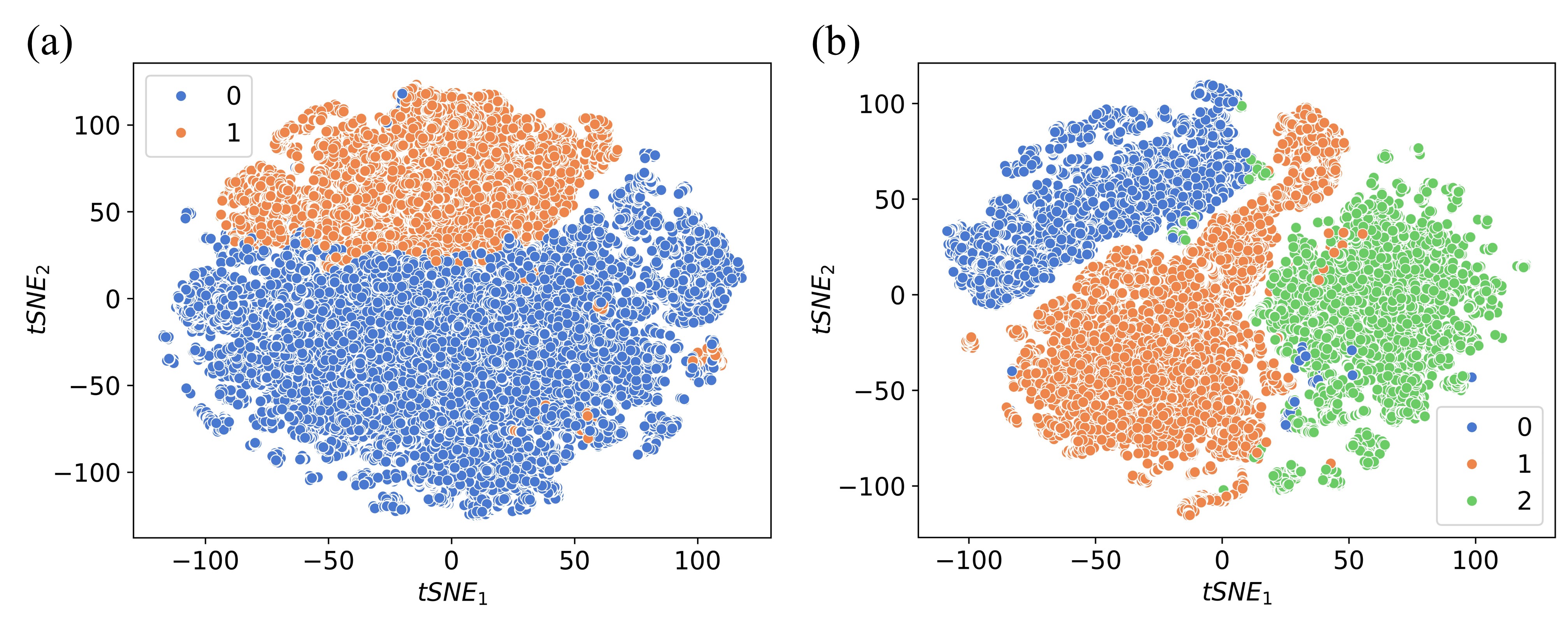}
\caption{\textbf{Sub-element classification}: t-SNE Plots for Li (a) and Mg (b) atoms extracted from atomic PDFs of COD database.}
\label{fig:subeltsne} 
\end{figure}

\begin{wrapfigure}{r}{0.6\linewidth}
    \begin{center}
        \includegraphics[width=0.99\linewidth]{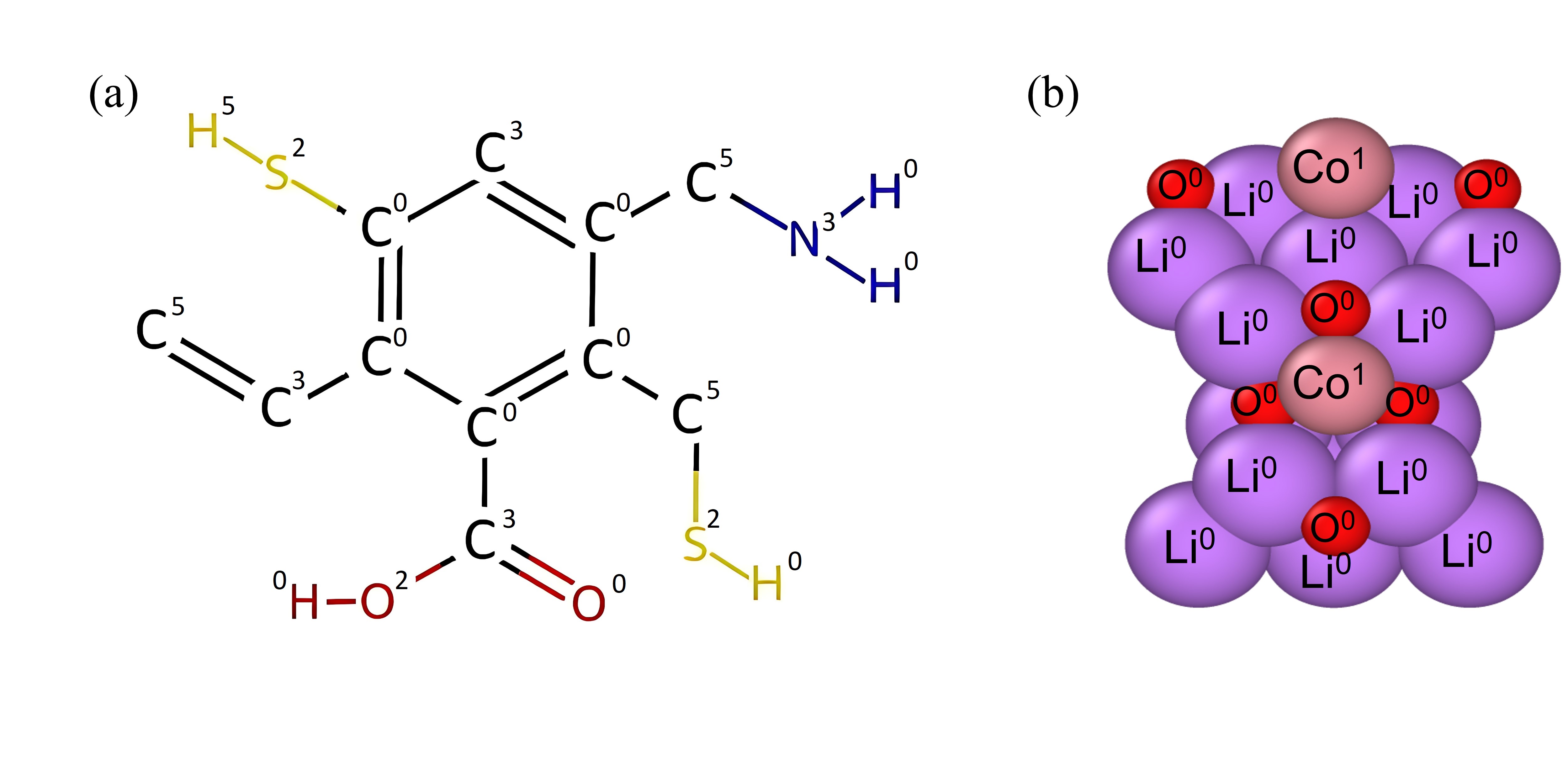}
    \end{center}    
\caption{\textbf{Examples illustrating the division of elements into sub-elements based on their environment}:  a hypothetical organic compound (a) and Li$_8$CoO$_6$ (b) crystal with ID mp-27920. The numbers at the top right of elements correspond to sub-element indexes.}
    \label{fig:subelement}
\end{wrapfigure}

To establish a universal criterion for categorizing these interactions, we divided the elements in our training dataset into “sub-elements” based on the local environment of atoms in compounds. We use the PDF and the oxidation state of each element to determine the number of possible sub-elements. It is crucial to note that information about the specific interaction type in molecular structures is often missing, which can lead to algorithmic errors. Recognizing that bond length contains information about interactions, we used unsupervised clustering to categorize elements. This approach is similar to the methods used manually in the development of classical force fields\cite{Damm1997OPLS}.
We evaluated several algorithms and ultimately selected the k-means (Km) algorithm due to its speed and simplicity. For fitting, we utilized structures from the Crystallography Open Database (COD) \cite{Gražulis2009Crystallography,Gražulis2012Crystallography}. Detailed information is provided in Appendix \ref{sec:appendixA}. To illustrate the workings of our approach, we present examples of sub-element classifications for lithium (Li) and magnesium (Mg) atoms in Fig. \ref{fig:subeltsne}. In these examples, Li and Mg atoms are classified into two and three groups, respectively. These t-SNE (t-distributed stochastic neighbor embedding) reductions of atomic PDFs provide a visualization of the principles behind atom division into sub-elements.
We fitted an individual model for each element in our dataset, resulting in a total of 96 models. The final size of the dictionary, including sub-element tokens, was ${V_{size}=565}$. Examples of such differentiation are presented in Fig. \ref{fig:subelement}.

\subsection{Datasets}
We trained our elEmBERT model to perform various classification tasks. To do this, we used the [CLS] token and added an additional layer to the BERT module with the same number of neurons as there are classes in the dataset. Our first task involved using the Materials Project (MP) metallicity dataset to predict the metallicity of materials based on structural information \cite{Jain2013Commentary,Ong2015Materials}. Next, we employed a portion of the datasets gathered for the CegaNN model \cite{Banik2023CEGANN}. This led us to undertake a classification task known as the Liquid-Amorphous (LA) task, which revolves around distinguishing between liquid and amorphous phases of silicon (Si). The LA dataset comprises 2,400 Si structures, evenly divided between amorphous and liquid phases (50\% each). Importantly, these Si structures lack symmetry and differ solely in terms of density and coordination number. In addition to these tasks, we evaluated the elEmBERT model's ability to classify material polymorphs across different dimensionalities, specifically clusters (0D), sheets (2D), and bulk structures (3D). Carbon, with its wide range of allotropes spanning these dimensionalities, served as an excellent system for assessing the efficiency of our network model in dimensionality classification (DIM task). The DIM dataset contained 1,827 configurations. Finally, we ventured into characterizing the space group of crystal structures, encompassing a total of 10,517 crystal structures distributed among eight distinct space groups (SG task) \cite{Ziletti2018Insightful}. 
Expanding beyond inorganic material datasets, we incorporated organic compounds, which greatly outnumber their inorganic counterparts. This expansion encompasses an extended range of properties, including biochemical and pharmaceutical aspects. To rigorously validate our model, we turned to benchmark datasets from MoleculeNet \cite{Wu2018MoleculeNet}, specifically BBBP (Blood-Brain Barrier Penetration), ClinTox (Clinical Toxicity), BACE ($\beta$-Secretase), SIDER (Side Effect Resource) \cite{Kuhn2016SIDER}, and Tox21. Notably, the Tox21 and SIDER datasets encompass 12 and 27 individual tasks, respectively, each corresponding to specific toxicity predictions. These datasets cover a diverse array of chemical compounds and provide a comprehensive assessment of our model's predictive performance for binary properties or activities associated with organic molecules. In this context, a positive instance signifies that a molecule possesses a specific property, while a negative instance indicates its absence. The MoleculeNet dataset primarily comprises organic molecules represented in Simplified Molecular Input Line Entry System (SMILES) format. For purposes of analysis, we converted these SMILES formulas into the standard XYZ format using the Open Babel software \cite{O'Boyle2011Open} and the RDKit package \cite{RDKit}. To evaluate our model's performance, we employed the "Receiver Operating Characteristic - Area Under the Curve" (ROC-AUC) metric, a common measure for assessing binary classification quality. ROC-AUC quantifies the model's ability to differentiate between positive and negative classes based on predicted probabilities.

\subsection{Training procedure}
In the following sections, we will present the results of prediction models with specific parameters, including an embedding size of 32, 2 attention heads, and 2 layers. These parameter choices have been identified as optimal across all datasets considered in this study. We explored two model versions, V0 (where the Km block is omitted) and V1, as discussed previously. These models were implemented using Keras \cite{Chollet2015Keras} and were trained with eight-fold splitting. This procedure is responsible for the splitting of the dataset and the initiation of model weights. We split the datasets into three subsets: the training set, the validation set, and the test set, with an 80:10:10 ratio. We employed the categorical crossentropy loss function with the Adam optimizer, setting the learning rate at 0.001. The batch size was set to 128 for the MP task and 32 for the other datasets. The number of epochs was two and four times the batch size, respectively.

\section{Results}
The ROC-AUC values reported in Table \ref{tab:benchmark1} represent averages over all test set prediction values. The table clearly demonstrates that the prediction efficiency improves as the number of tokens increases, particularly for inorganic compounds. In the LA task, using single-element inputs results in only 50\% accuracy, which is comparable to random guessing. However, incorporating sub-elements significantly enhances the performance, leading to an impressive ROC-AUC of 0.983. Our approach also demonstrates improved scores across other datasets. In the subsequent sections, we will delve into each dataset, from Matbench to Tox21, and examine the elEmBERT-V1 model in more detail, providing comprehensive insights into the predictions. 

\begin{table}[h!]
\caption{Performance of elEmBERT models applied to datasets used in this work. A \textbf{bold} font indicates the best performance, an  \underline{underline} represents the second-best performance, and the last column presents previous results obtained from other models. V0 represents models that use chemical element embeddings, while V1 uses sub-element embeddings as input for the BERT module.}
\begin{center}
\begin{tabular}{lcccc}
\hline\hline
\textbf{Benchmark} & \textbf{V0} & \textbf{V1} & \textbf{Best} \\
\hline
MP metallicity &  \underline{0.961} & \textbf{0.965}  & 0.950 \cite{Chen2021AtomSets} \\
SG & 0.945  & \underline{0.952} & \textbf{1.000} \cite{Banik2023CEGANN}\\
LA & 0.500  & \underline{0.983}  & \textbf{1.000} \cite{Banik2023CEGANN}\\
DIM & 0.866  & \underline{0.958}  & \textbf{1.000} \cite{Banik2023CEGANN}\\
BACE & 0.732  & \underline{0.789}  & \textbf{0.888} \cite{Li2022GLAM}\\
BBBP & 0.903  & \underline{0.909} & \textbf{0.932} \cite{Li2022GLAM}\\
ClinTox & \textbf{0.962}  & \underline{0.959}  & 0.948 \cite{{Li2021TrimNet}}\\
HIV & \textbf{0.982}  & \underline{0.972} & 0.776 \cite{Baek2021Accurate}\\
SIDER & \textbf{0.778} & \underline{0.773}  & 0.659 \cite{Li2022GLAM}\\
Tox21 & \underline{0.965}  & \textbf{0.967} & 0.860 \cite{{Li2021TrimNet}}\\
\hline
\end{tabular}
\end{center}
\label{tab:benchmark1}
\end{table}

\subsection{MP metallicity}
\begin{figure*}[h]
\centering
\includegraphics[width=\linewidth]{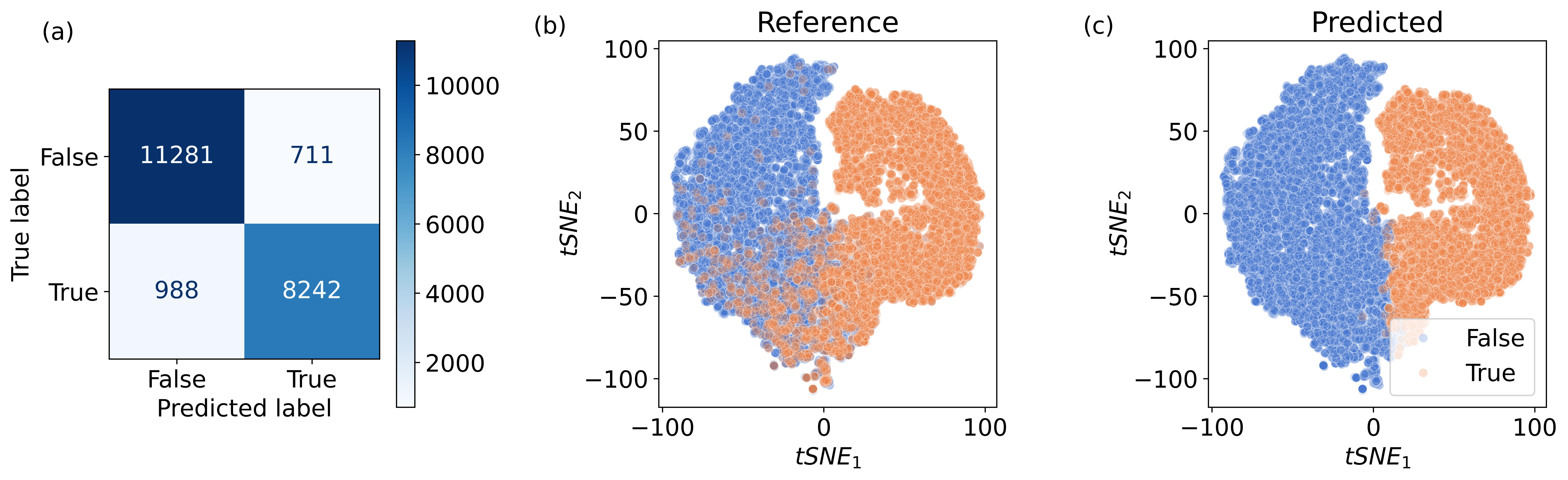}
\caption{\textbf{MP metallicity}: Confusion matrix (a) and visualization of [CLS] token embeddings for the MP metallicity dataset for the reference (b) and predicted (c) datasets: blue circles denote negative labels (not metal) and orange dots represent positive labels (metal).}
\label{fig:matbench} 
\end{figure*}

In this task, the objective is to predict whether a material is metallic or non-metallic, a property initially determined by electronic structure calculations. Fig. \ref{fig:matbench}a illustrates the confusion matrix and presents the performance of the elEmBERT-V1 model in classifying MP metallicity. The dataset for this task comprises 106,113 samples of training structures and 21,222 samples of test structures. Our trained model achieves a binary accuracy of approximately 0.910 and an AUC of 0.965 on the test set. 
In Fig. \ref{fig:matbench}b, the t-SNE plot shows the embeddings of the entire reference dataset, categorized by labels. A smooth differentiation among labels within the feature space is revealed. Fig. \ref{fig:matbench}c illustrates how the reference dataset is classified by our model. The classification layer creates a clear separation in the feature space, in contrast to the diffuse boundary in the reference dataset. Primary errors are located at the boundary, where the model sometimes struggles to effectively capture diffusive behavior. The metallicity prediction task highlights elEmBERT's remarkable capability to characterize these binary properties of crystals. The results exceed the capabilities of previously published models, including those of GNNs.

\subsection{LA, DIM and SG datasets}

\begin{figure*}[h]
\centering
\includegraphics[width=\linewidth]{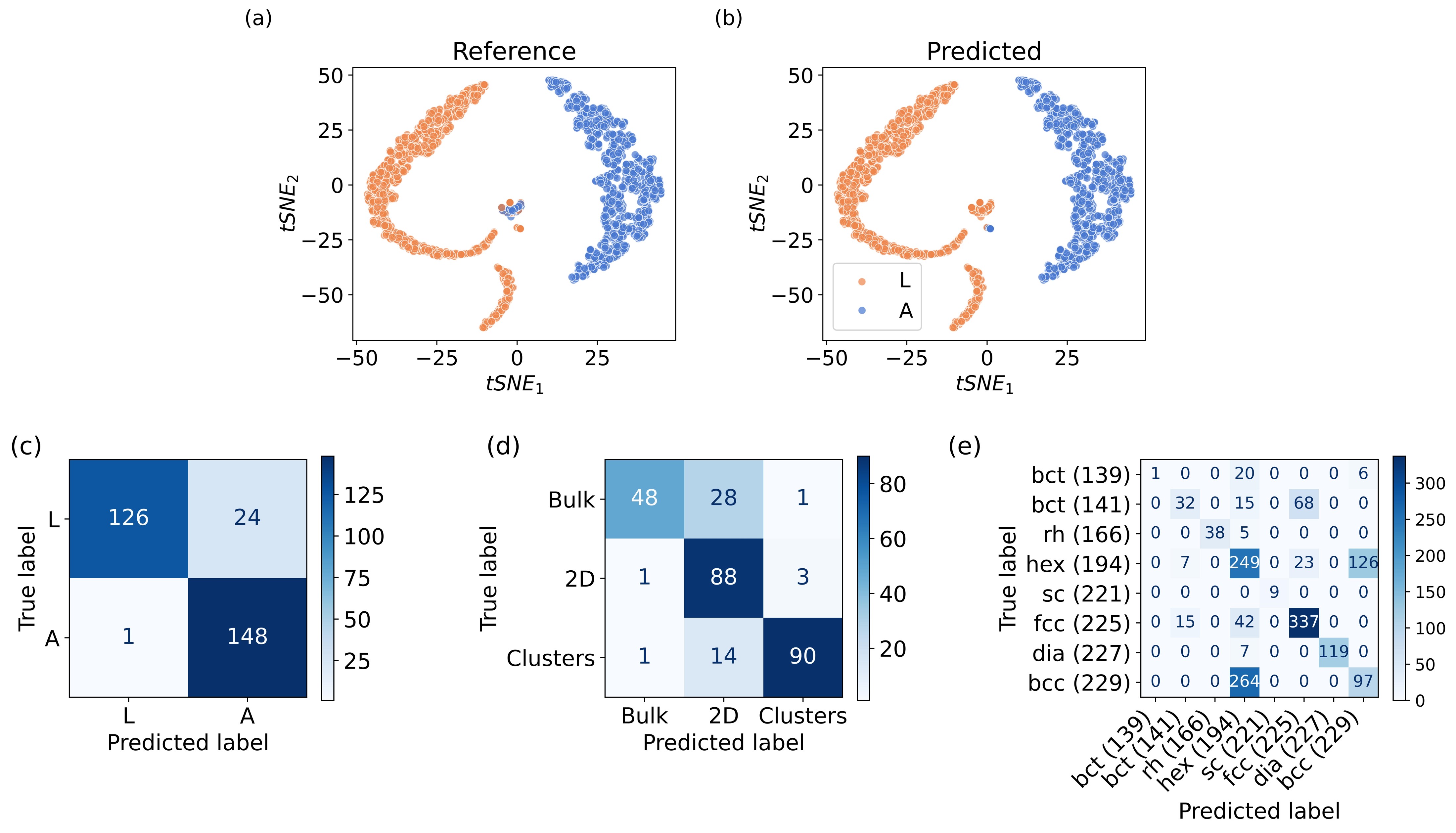}
\caption{\textbf{Classification task of inorganic compunds}. Top row: Visualization of [CLS] Token Embeddings for the LA Dataset: a) reference labels and b) predicted labels. The embeddings are represented using blue circles for liquid phase labels and orange dots for amorphous labels. Bottom row: Confusion matrix analysis of the LA (c), DIM (d), and SG (e) datasets.}
\label{fig_lasgd} 
\end{figure*}

This section presents the results from benchmarks conducted on the CegaNN dataset, with an initial focus on the LA classification task. Figs. \ref{fig_lasgd}a and \ref{fig_lasgd}b show the embedding representation of Si structures based on their labels, reduced through the t-SNE algorithm. Our model effectively separates the structures into distinct clusters, with two clusters clearly corresponding to their respective classes. However, one cluster exhibits intermixing of structures, which challenges accurate recognition by the model.
The confusion matrices shown in Fig. \ref{fig_lasgd}c-e provide insights into the performance of the elEmBERT-V1 model across the LA, DIM, and SG datasets. The model achieves a high ROC-AUC of approximately 0.958 on the DIM task's test set and a slightly higher value of 0.968 on the SG dataset. These confusion matrices illustrate the model's ability to identify and categorize each structure accurately. It is worth noting that the model faces challenges in distinguishing the bcc (229) structure from others in the SG dataset. This challenge arises from the structural similarities between the bcc structure and others, resulting in identical geometrical representations unless the orientational order of the particles is considered.
While the CegaNN model achieves approximately 100\% accuracy in this benchmark, our model approaches this level of performance, also exhibiting strengths in versatility, speed, and simplicity.

\subsection{Tox21 dataset}

\begin{figure*}[h!]
\centering
\includegraphics[width=\linewidth]{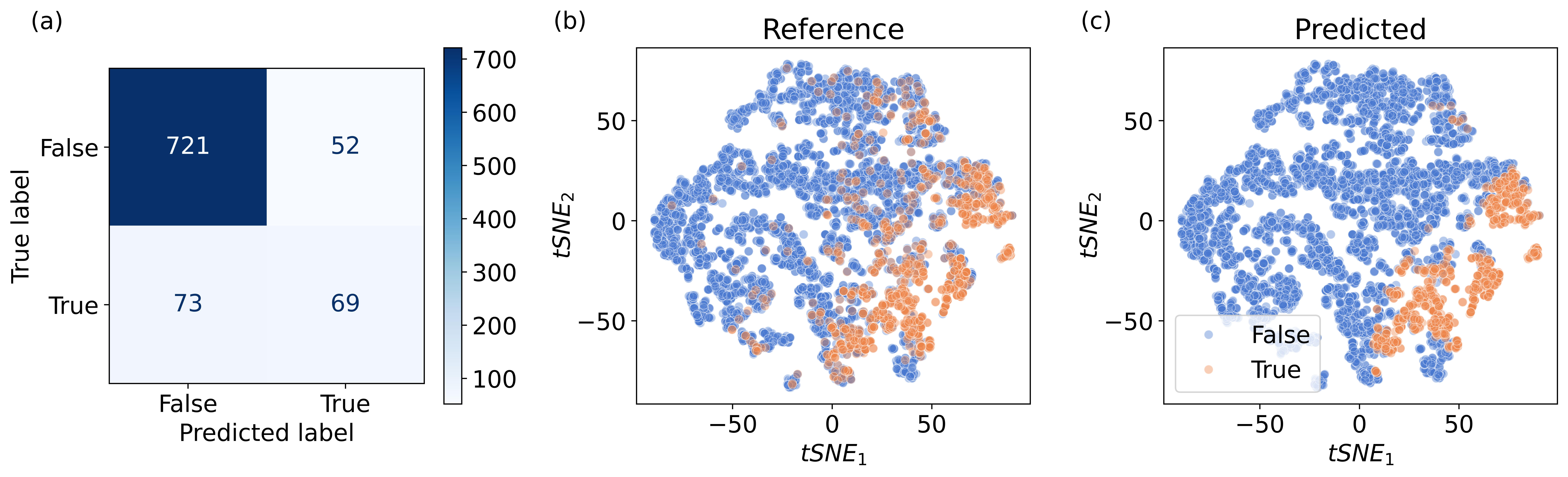}
\caption{\textbf{Classification results of sr-mmp task from Tox21 dataset}: a) Confusion matrix of predicted labels on the test set. b) t-SNE feature representation of the entire reference dataset according to their labels. c) Feature representation of the predicted labels.}
\label{fig:tox21} 
\end{figure*}

The Tox21 dataset is a collection of chemical compounds evaluated for their toxicity against 12 biological targets. With over 8,000 compounds, it serves as a valuable resource for predicting the toxicity and potential adverse effects of various chemical compounds. Our model, trained on the Tox21 dataset, demonstrated impressive performance, achieving an average AUC of 0.96 across all 12 toxicity prediction tasks\cite{Wu2018MoleculeNet}. The results of these individual tasks are presented in Table \ref{tab:tox21}, enabling a comprehensive evaluation of the model's performance on each toxicity task.\\
Comparing our results with those of the Meta-Molecular GNN (MMGNN) model\cite{Guo2021Few-shot} highlights the significant advantages of our approach. Fig. \ref{fig:tox21} shows the confusion matrix of the test set and the t-SNE projection representing the features of the sr-mmp task in the Tox21 dataset. As shown, our model predicts distinct patterns in the t-SNE projections, with each label value occupying a specific region (Fig. \ref{fig:tox21}b). The molecular embedding visualizations are also available in the MMGNN model report for the sr-mmp task\cite{Guo2021Few-shot}. In contrast, our feature space exhibits more structure, with positive values being less dispersed across all compounds. Our model primarily has a few points that are significantly distant from the positive value region. Both elEmBERT models successfully identify the boundary between these two classes and make predictions (Fig. \ref{fig:tox21}). Errors primarily arise from diffuse boundary regions and points located far from the true cluster. This observation holds true for all Tox21 tasks.
Similarly, prediction analysis for the BACE, BBBP, ClinTox, HIV, and SIDER datasets is detailed in Appendix \ref{sec:appendixB}. 

\begin{table*}[h!]
    \centering
    \caption{ROC-AUC performance of different tasks from the Tox21 dataset. MMGNN denotes the prior top-performing results\cite{Guo2021Few-shot}. Averaged score across all tasks showed in Table \ref{tab:benchmark1}.}
    \begin{tabular}[width\linewidth]{ccccccc}
        \hline
        \hline
        \textbf{Model} & \textbf{nr-ahr} & \textbf{nr-ar-lbd} & \textbf{nr-arom} & \textbf{nr-ar} & \textbf{nr-er-lbd} & \textbf{nr-er} \\
        \hline
        V0 & 0.955 & 0.987 & 0.980 & 0.983 & 0.978 & 0.932 \\
        V1 & 0.961 & 0.989 & 0.979 & 0.982 & 0.978 & 0.935 \\
        MMGNN & - & - & - & - & - & -  \\
        \hline
         & \textbf{nr-ppar-gamma} & \textbf{sr-are} & \textbf{sr-atad5} & \textbf{sr-hse} & \textbf{sr-mmp} & \textbf{sr-p53}  \\
        V0 & 0.988 & 0.913 & 0.985 & 0.974 & 0.938 & 0.972 \\
        V1 & 0.986 & 0.913 & 0.983 & 0.974 & 0.946 & 0.973 \\
        MMGNN  & - & - & - & 0.748 & 0.804 & 0.790 \\
        
        \hline
    \end{tabular}
    \label{tab:tox21}
\end{table*}

\section{Discussion}
One key aspect contributing to the efficacy of the elEmBERT model is its ability to capture the complex geometric attributes inherent in chemical structures, enabling accurate prediction of multiple chemical properties. Our elEmBERT-V0 model is similar to previous proven models such as MolBERT \cite{Fabian2020Molecular} or MolBART\cite{Irwin2022Chemformer}. These models utilize SMILES as input data for molecular representations. However, in chemistry-related tasks, it is important to consider information about the spatial distribution of atoms within compounds. Traditional text-based (composition or SMILES formats) models are inherently limited in capturing important structural and conformational details of compounds. This limitation is a significant drawback of such models. Our elEmBERT-V1 model aims to address and mitigate this drawback.
\begin{figure*}[h!]
\centering
\includegraphics[width=\linewidth]{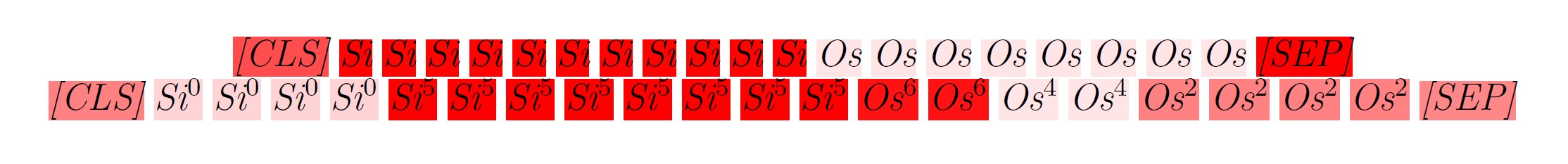}
\caption{\textbf{Visualization of the attention scores from elEmBERT}. The top row displays the output from V0 and the bottom row the one from the V1 model.}
\label{fig:attention} 
\end{figure*}
The encoding of atomic PDFs into sub-element tokens and using their corresponding embedding vectors as inputs to the neural network ensures a comprehensive consideration of both local atomic and global compound features.This approach is particularly significant in the context of "structure to property" tasks, where a thorough understanding of the atomic environment is crucial for accurate prediction of compound properties. As has been shown in the LA task, the V1 model excels at capturing the geometric nuances inherent in chemical structures, an observation that extends to other inorganic datasets. For instance, we can consider the visualization of the attention scores of the last layer from the V0 and V1 models. Fig. \ref{fig:attention} demonstrates color-coded values of attention scores for materials containing Si and Osmium (Os) atoms. Notably, the V0 model assigns identical values for Si and Os atoms. In contrast, the sub-element approach divides Si and Os atoms into two (Si$^0$, Si$^5$) and three subgroups (Os$^2$, Os$^4$, Os$^6$) respectively. Following this, the V1 model assigns diverse attention scores to individual sub-element tokens, thereby expanding the model's parameters and enhancing its predictive performance. \\
Despite demonstrating sufficient performance on inorganic datasets, our model shows a predictive gap compared to other models on the drug-related BACE and BBBP datasets. This discrepancy can be attributed to conformational variations among compounds present in these datasets. While we transformed SMILES into 3D structures, the existence of multiple conformers for a given SMILES formula introduces complexities. Not all conformers generated in this study may precisely match the actual structural configuration, thus diminishing the predictive power of the model. \\
These issues may also explain the minor prediction differences observed between the V0 and V1 models in organic benchmarks. To address these issues, using accurately and thoroughly described structures could prove advantageous. Additionally, increasing the number of sub-elements may more effectively capture the complexity of chemical structures. Incorporating positional information into uniformly described databases could further enhance the predictive accuracy of our workflow. Comparative examples are provided in Appendix \ref{sec:appendixC}. Moreover, the elEmBERT architecture supports a pre-training strategy on a large structural dataset, enabling the model to capture various chemical differences and apply this knowledge to smaller datasets.
\section{Conclusions}
In conclusion, the deep learning model presented in this paper signifies a marked advancement in the application of machine learning to computational chemistry. By integrating the attention mechanism and a transformer-based approach, our model can capture both local and global properties of chemical compounds, enabling highly accurate predictions of chemical properties that outperform similar approaches. The combination of principal component analysis and k-means clustering for sub-elements accounts for the nuanced effects stemming from electronic structure, a fact confirmed through the analysis of numerous chemical databases. Our classification approach, which relies on compound embeddings, results in substantially improved prediction accuracy compared to previously published scores. Additionally, t-SNE projections provide valuable insights into the classification mechanisms and can pinpoint sources of erroneous predictions. Future enhancements can likely be attained through augmentation of the sub-elements in the elEmBERT-V1 model and by the use of more sophisticated atomic descriptors.
\section{Acknowledgments}
The work has partially been carried out within the framework of the EUROfusion Consortium and received funding from the Euratom research and training programme by Grant Agreement No. 101052200-EUROfusion. The views and opinions expressed herein do not necessarily reflect those of the European Commission. The computational results have been obtained using the HPC infrastructure LEO of the University of Innsbruck.

\section{Availability}
The source code, trained weights, and example notebooks of elEmBERT available at \href{https://github.com/dmamur/elembert}{https://github.com/dmamur/elembert}.\\
All data used in this paper are publicly available and can be accessed from various sources. The structure files for the MP metallicity dataset are accessible at \href{https://matbench.materialsproject.org/}{https://matbench.materialsproject.org/}. The LA, SG, and DIM datasets are available at 
\href{https://github.com/sbanik2/CEGANN/tree/main/pretrained}{https://github.com/sbanik2/CEGANN/tree/main/pretrained}. The BACE, BBBP, ClinTox, HIV, and SIDER datasets can be retrieved from \href{https://moleculenet.org/}{https://moleculenet.org/}. Structure files for the Tox21 dataset can be obtained from \href{https://tripod.nih.gov/tox21/challenge/data.jsp}{https://tripod.nih.gov/tox21/challenge/data.jsp}. 

\begin{appendices}
\section{Sub-element approach}
\label{sec:appendixA}
\begin{wrapfigure}{r}{0.4\linewidth}
\vspace{-6mm}
    \begin{center}
        \includegraphics[width=0.99\linewidth]{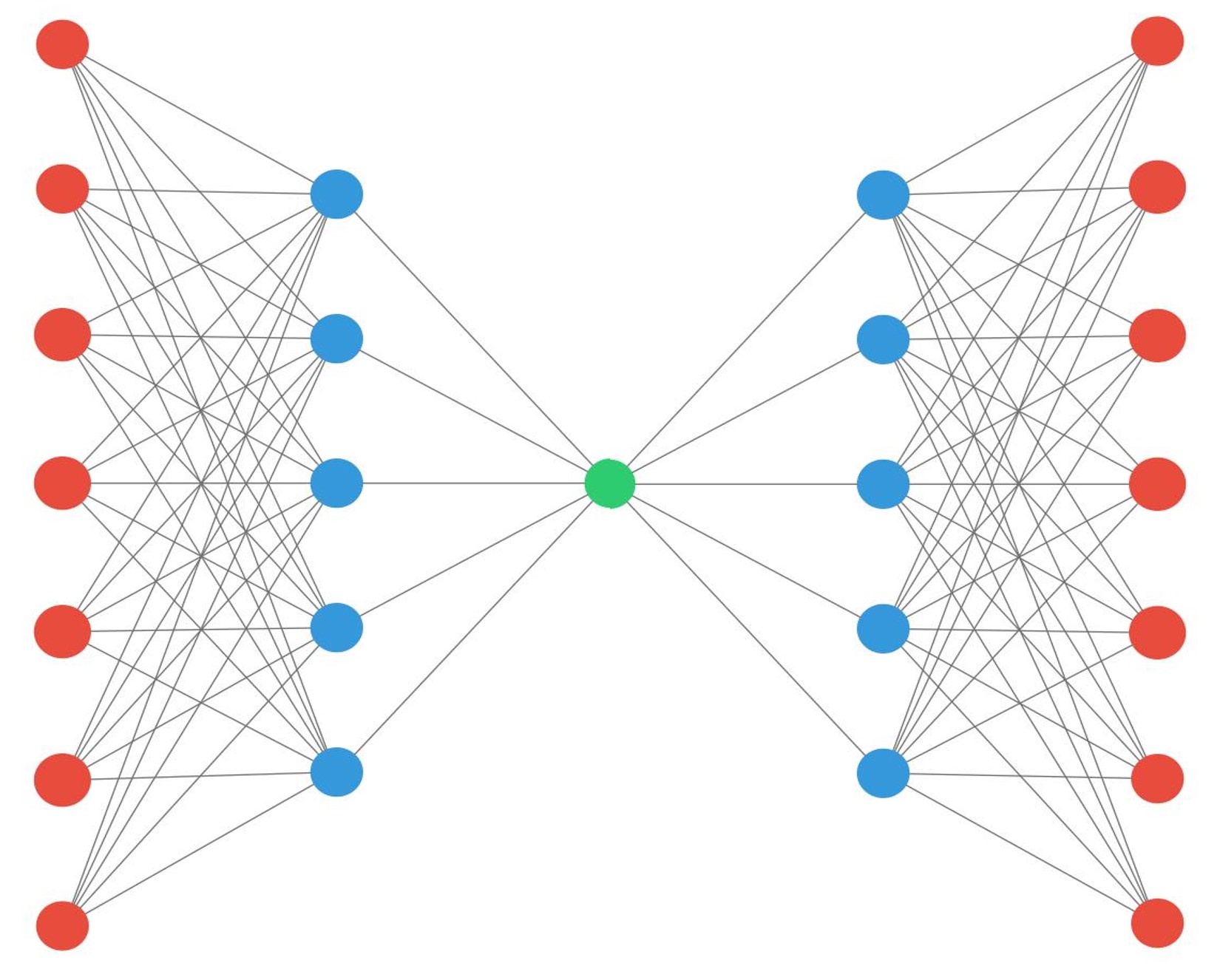}
    \end{center}
    
\caption{Neural network diagram for encoding-decoding approach.}
    \label{fig:nn}
\end{wrapfigure}

\begin{figure}[h!]
\centering
\includegraphics[width=0.6\linewidth]{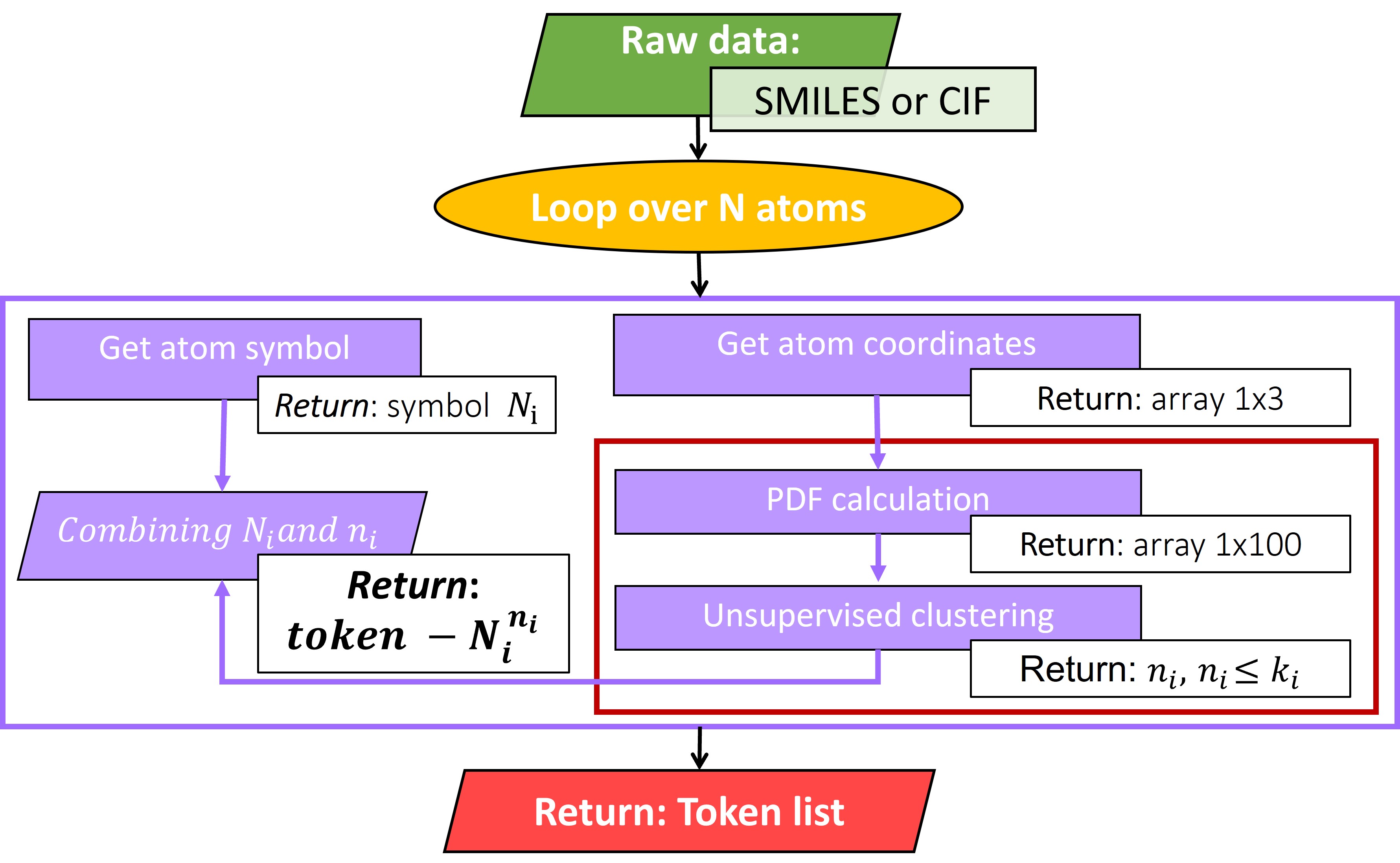}
\caption{\textbf{Workflow of the unsupervised classification module}. The process starts with compounds provided in Crystallographic Information File (CIF) or SMILES format as input data. The module first extracts the atomic symbols and Cartesian coordinates of all atoms from this input. Next, PDFs are computed based on these coordinates. The computed PDFs are then subjected to unsupervised clustering. The resulting clusters are mapped to the atomic symbols, facilitating their conversion into tokens for further analysis.}
\label{fig:scheme} 
\end{figure}

\begin{figure}[h]
\centering
\includegraphics[width=0.67\linewidth]{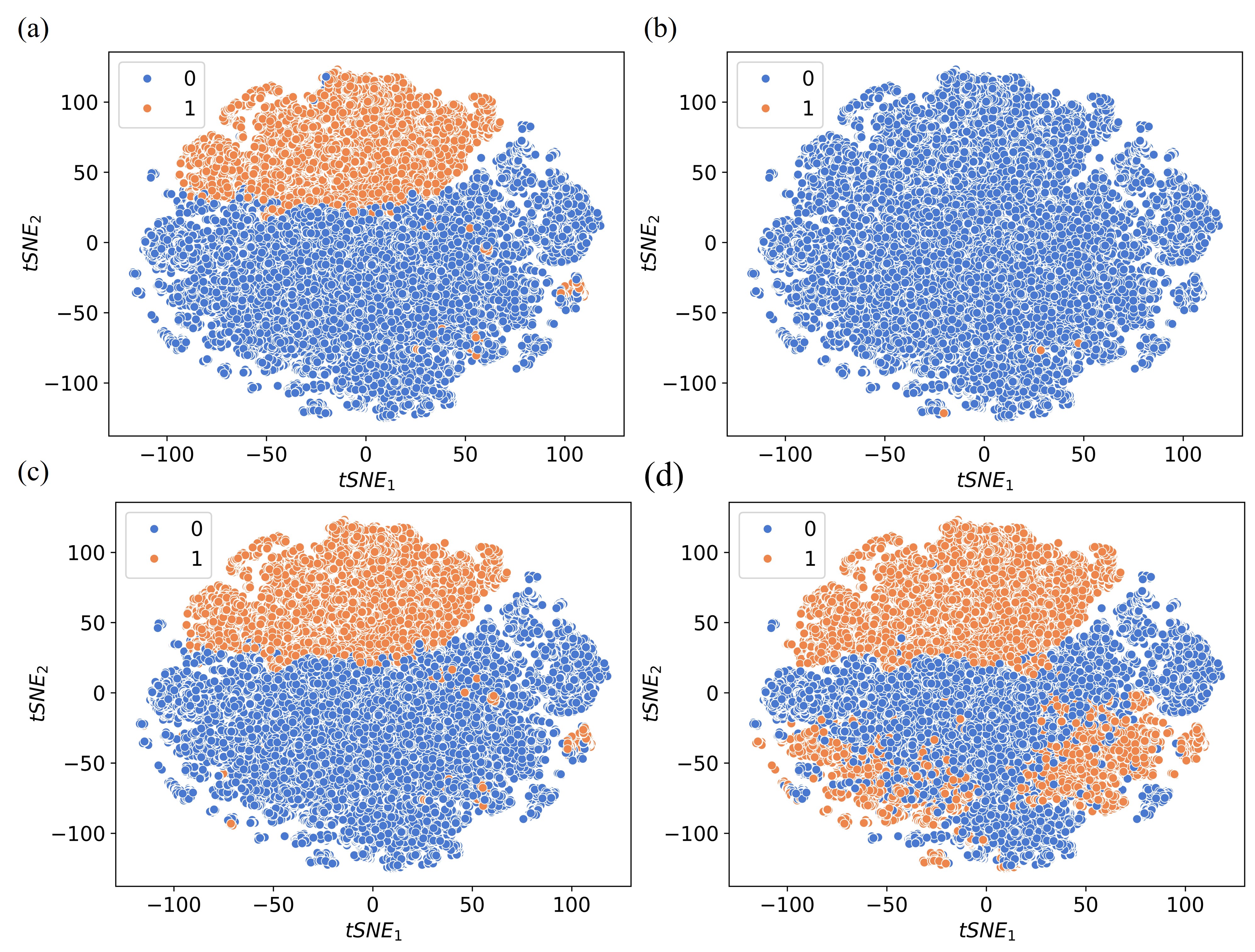}
\caption{\textbf{Sub-element classification of Litium atoms}. t-SNE Plots for Mg for k-means (a), Feature Agglomeration (b), NN encoder-decoder (c) and PCA-k-means (d) algorithms. }
\label{fig:cluLi} 
\end{figure}

\begin{figure}[h]
\centering
\includegraphics[width=0.7\linewidth]{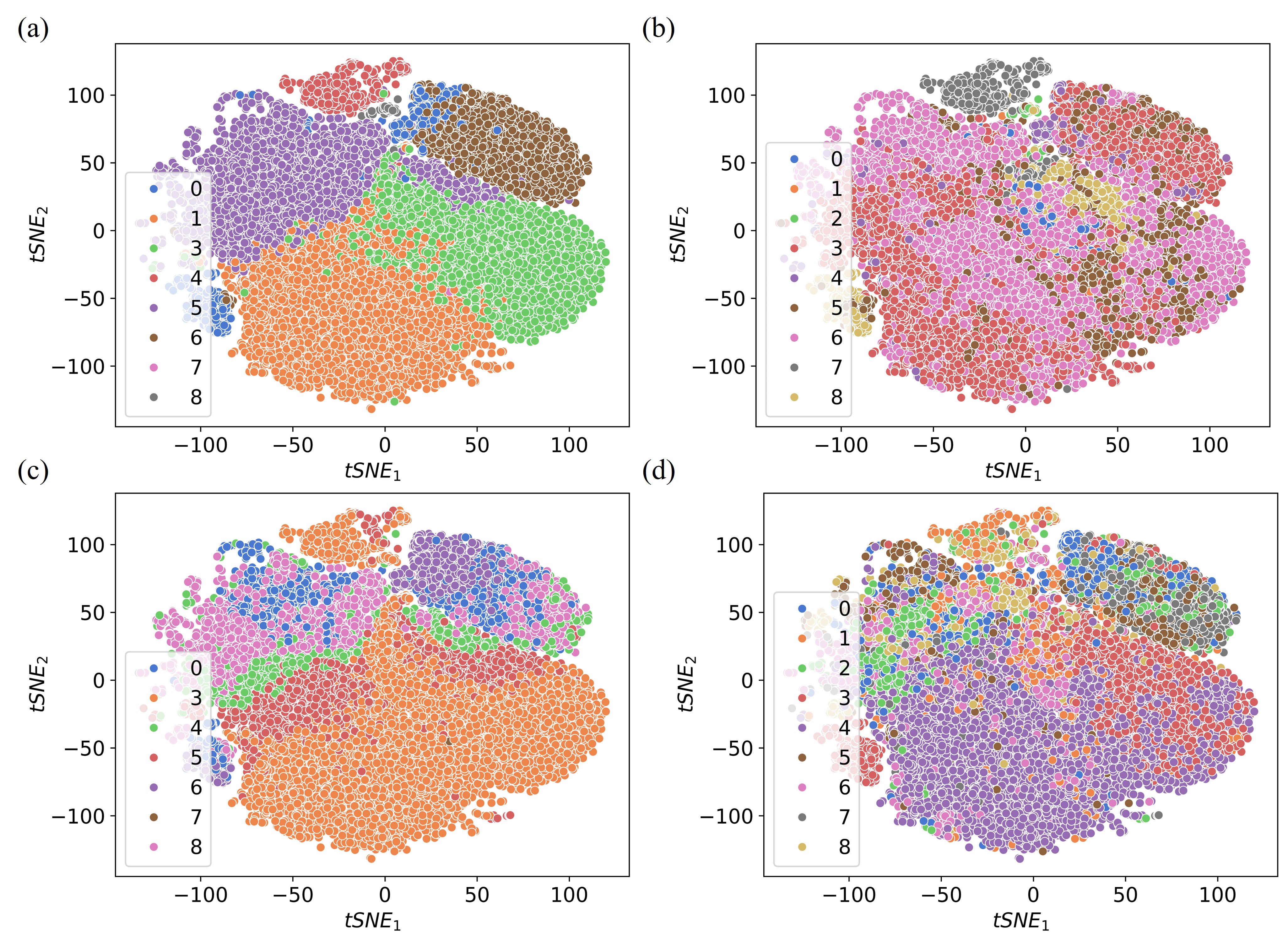}
\caption{\textbf{Sub-element classification of Chlorine  atoms}. t-SNE Plots for Li for k-means (a), Feature Agglomeration (b), NN encoder-decoder (c), and PCA- k-means (d) algorithms. }
\label{fig:cluCl} 
\end{figure}

To categorize elements into sub-elements (Fig. \ref{fig:scheme}), we examined various unsupervised classification algorithms, including k-means, Feature Agglomeration, neural network encoder-decoder models, and Principal Component Analysis (PCA) of PDFs combined with the k-means algorithm. The number of clusters was determined based on the oxidation states of the elements, as detailed in Table \ref{tab:elements}. 
\\
Clustering was performed using structures from the Crystallography Open Database with the scikit-learn package \cite{Pedregosa2011Scikit-learn}, with the exception of the NN encoder-decoder models. The NN architecture was manually created and symmetric, designed to reproduce the input data through a series of progressively reduced layers (Fig. \ref{fig:nn}). The central layer is configured with a number of neurons corresponding to the number of clusters. As the input atom PDF traverses the network, the model aims to identify the neuron in this central layer that produces the highest activation value. This process allows the model to determine the cluster to which the input atom is assigned. \\ 
The sub-element approach involves calculating and extracting atomic PDFs from all structures in the database. These PDFs were then aggregated according to atomic type to form the input data arrays. The models obtained from this clustering process were subsequently integrated into the elEmBERT model or could be applied as a preprocessing step. Benchmark values for different classification methods are presented in Table \ref{tab:nnbenchmark}, while visual differences in predicting sub-element indices among these methods are illustrated in Fig. \ref{fig:cluLi} and \ref{fig:cluCl}.
\begin{table}[h!]
\centering
\caption{Cluster numbers (k) for clustering selected based on element (El) oxidation states.}
\begin{tabular}[width=\linewidth]{|ccc|ccc|ccc|ccc|ccc|ccc|}
\hline
\textbf{N} & \textbf{El} & \textbf{k} & \textbf{N} & \textbf{El} & \textbf{k} & \textbf{N} & \textbf{El} & \textbf{k} & \textbf{N} & \textbf{El} & \textbf{k} & \textbf{N} & \textbf{El} & \textbf{k} & \textbf{N} & \textbf{El} & \textbf{k} \\
\hline
1 & H & 3 & 17 & Cl & 9 & 33 & As & 5 & 49 & In & 4 & 65 & Tb & 4 & 81 & Tl & 3 \\
2 & He & 2 & 18 & Ar & 2 & 34 & Se & 6 & 50 & Sn & 4 & 66 & Dy & 3 & 82 & Pb & 4 \\
3 & Li & 2 & 19 & K & 3 & 35 & Br & 7 & 51 & Sb & 4 & 67 & Ho & 2 & 83 & Bi & 4 \\
4 & Be & 3 & 20 & Ca & 3 & 36 & Kr & 2 & 52 & Te & 6 & 68 & Er & 2 & 84 & Po & 5 \\
5 & B & 4 & 21 & Sc & 4 & 37 & Rb & 3 & 53 & I & 7 & 69 & Tm & 3 & 85 & Ra & 1 \\
6 & C & 9 & 22 & Ti & 5 & 38 & Sr & 3 & 54 & Xe & 5 & 70 & Yb & 3 & 86 & Ac & 3 \\
7 & N & 9 & 23 & V & 7 & 39 & Y & 4 & 55 & Cs & 3 & 71 & Lu & 2 & 87 & Th & 4 \\
8 & O & 5 & 24 & Cr & 9 & 40 & Zr & 5 & 56 & Ba & 2 & 72 & Hf & 4 & 88 & Pa & 5 \\
9 & F & 2 & 25 & Mn & 11 & 41 & Nb & 6 & 57 & La & 3 & 73 & Ta & 6 & 89 & U & 6 \\
10 & Ne & 2 & 26 & Fe & 9 & 42 & Mo & 9 & 58 & Ce & 4 & 74 & W & 9 & 90 & Np & 6 \\
11 & Na & 3 & 27 & Co & 7 & 43 & Tc & 10 & 59 & Pr & 4 & 75 & Re & 10 & 91 & Pu & 7 \\
12 & Mg & 3 & 28 & Ni & 6 & 44 & Ru & 10 & 60 & Nd & 3 & 76 & Os & 10 & 92 & Am & 6 \\
13 & Al & 4 & 29 & Cu & 5 & 45 & Rh & 8 & 61 & Pm & 2 & 77 & Ir & 11 & 93 & Cm & 3 \\
14 & Si & 9 & 30 & Zn & 3 & 46 & Pd & 3 & 62 & Sm & 3 & 78 & Pt & 5 & 94 & Bk & 1 \\
15 & P & 9 & 31 & Ga & 4 & 47 & Ag & 5 & 63 & Eu & 3 & 79 & Au & 6 & 95 & Cf & 1 \\
16 & S & 9 & 32 & Ge & 6 & 48 & Cd & 3 & 64 & Gd & 4 & 80 & Hg & 4 & 96 & Rn & 1 \\
\hline
\end{tabular}
\label{tab:elements}
\end{table}

\begin{table}[h!]
\centering
\caption{Performance of different classification models applied to datasets used in this work.}
\begin{tabular}{lcccc}
\hline
Benchmark        & Km   & NN   & FA   & PCA  \\
\hline
MP metallicity   & 0.967 & 0.969 & 0.967 & 0.967 \\
SG               & 0.954 & 0.954 & 0.953 & 0.967 \\
LA               & 0.987 & 0.988 & 0.988 & 0.987 \\
DIM              & 0.881 & 0.948 & 0.950 & 0.932 \\
BACE             & 0.871 & 0.856 & 0.817 & 0.845 \\
BBBP             & 0.887 & 0.893 & 0.924 & 0.896 \\
ClinTox          & 0.973 & 0.969 & 0.970 & 0.967 \\
HIV              & 0.978 & 0.979 & 0.977 & 0.977 \\
\hline
\end{tabular}
\label{tab:nnbenchmark}
\end{table}


\section{Organic benchmarks}
\label{sec:appendixB}

\subsection{BACE Dataset}
\begin{figure*}[h!]
\centering
\includegraphics[width=\linewidth]{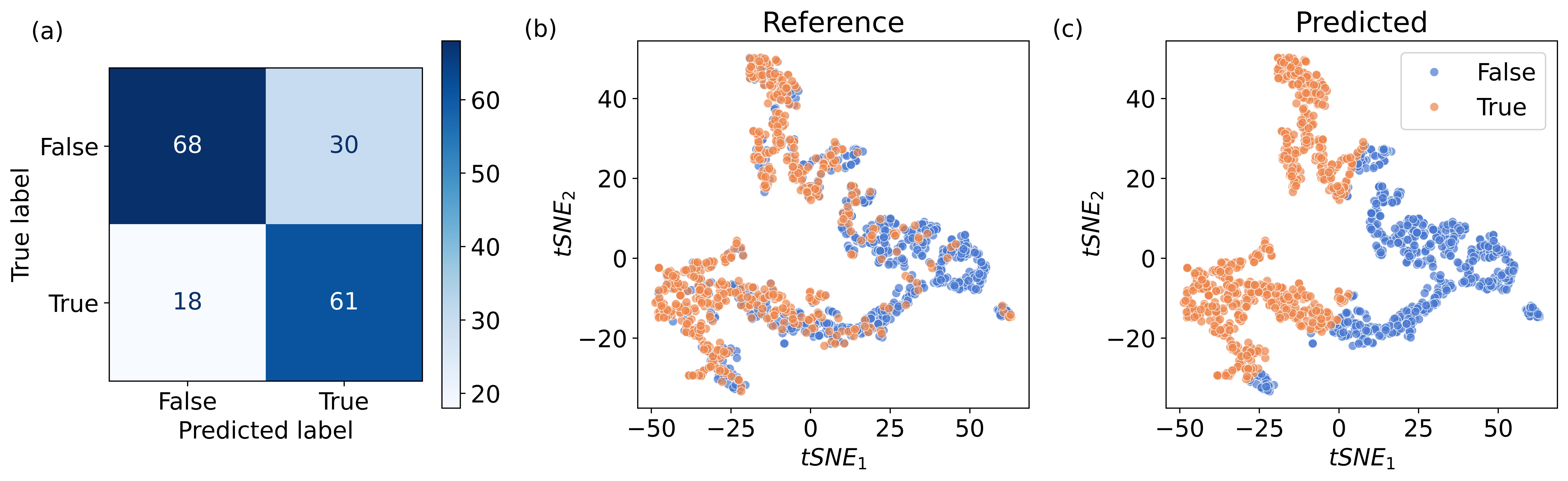}
\caption{\textbf{Classification of BACE data}: a) Confusion matrix of predicted labels on the test set. b) t-SNE feature representation of the entire reference dataset according to their labels. c) Feature representation of the predicted labels. }
\label{fig:bace} 
\end{figure*}
The BACE dataset includes 1,513 compounds classified as active or inactive inhibitors of the $\beta$-secretase enzyme, associated with Alzheimer's disease. Fig. \ref{fig:bace} shows the elEmBERT-V1 predictions, revealing two clusters with intermingled labels, which lead to some misclassifications.

\subsection{BBBP Dataset}
\begin{figure*}[t]
\centering
\includegraphics[width=\linewidth]{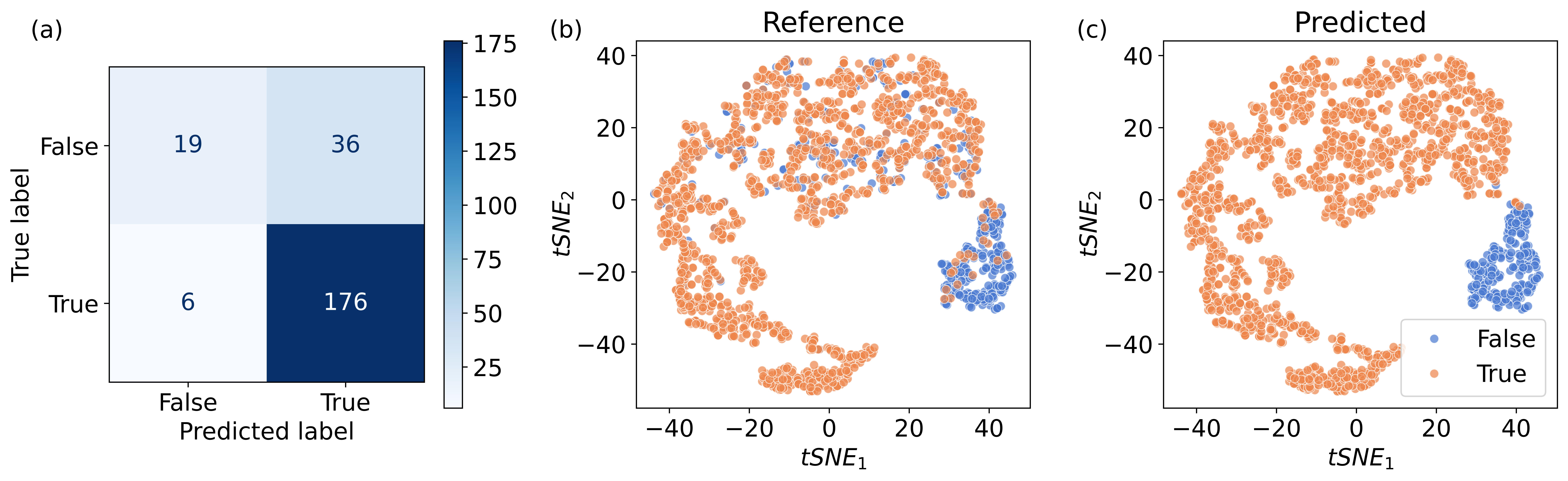}
\caption{\textbf{Classification of BBBP data}: a) Confusion matrix of predicted labels on the test set. b) t-SNE feature representation of the entire reference dataset according to their labels. c) Feature representation of the predicted labels. }
\label{fig:bbbp} 
\end{figure*}

The BBBP dataset contains 2,039 compounds, categorized by their ability to penetrate the blood-brain barrier. Fig. \ref{fig:bbbp} shows the confusion matrix and t-SNE plots, illustrating label separation.

\subsection{ClinTox Dataset}
\begin{figure*}[t]
\centering
\includegraphics[width=\linewidth]{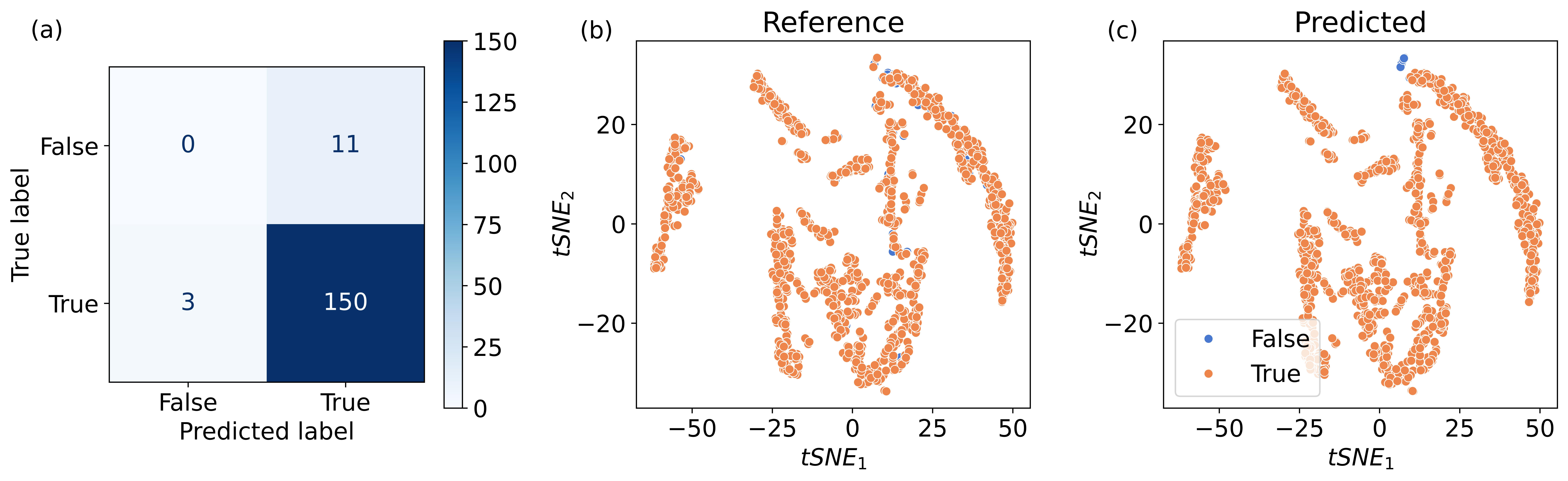}
\caption{\textbf{Classification of ClinTox FDA approval task}: a) The confusion matrix of predicted labels on the test set. b) The t-SNE feature representation of the entire reference dataset according to their labels. c) The feature representation of the predicted labels.}
\label{fig:clintox} 
\end{figure*}

Containing 1,491 compounds, the ClinTox dataset assesses clinical toxicity and FDA approval status. Fig. \ref{fig:clintox} shows the results. Due to the low number of negative instances, the model accurately predicts all negative values in a small region and only for the training set. 

\subsection{HIV Dataset}
\begin{figure*}[h]
\centering
\includegraphics[width=\linewidth]{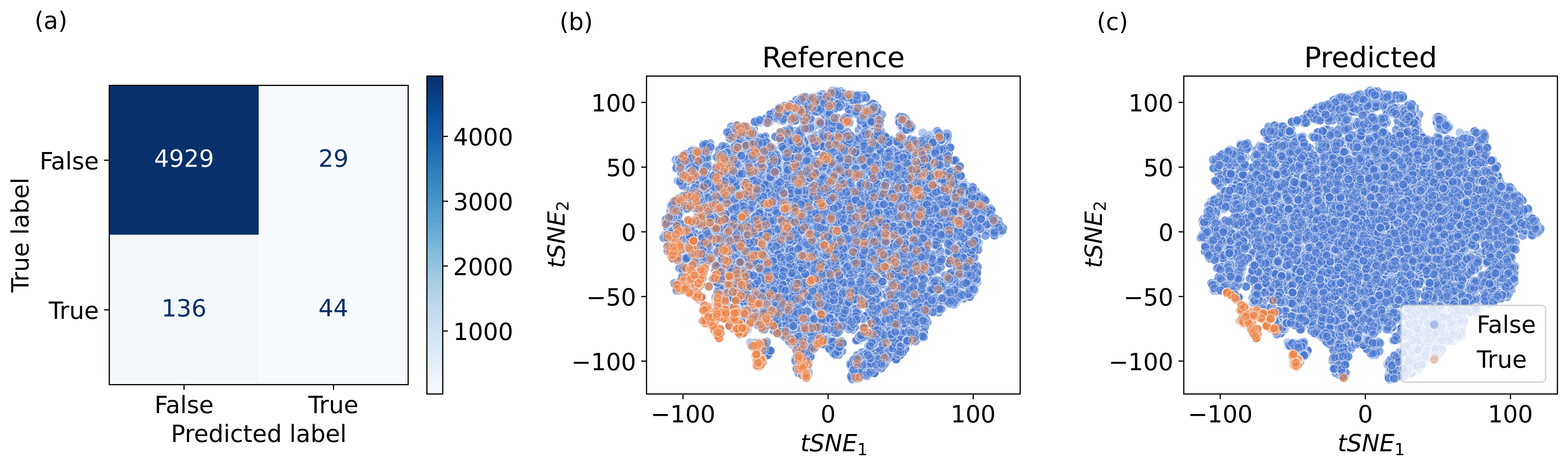}
\caption{\textbf{Classification of HIV data}: a) Confusion matrix of predicted labels on the test set. b) t-SNE feature representation of the entire reference dataset according to their labels. c) Feature representation of the predicted labels. }
\label{fig:hiv} 
\end{figure*}

The HIV dataset, which includes approximately 41,000 data points with 1,443 positives, is used to predict patient treatment responses and drug resistance. Fig. \ref{fig:hiv} highlights the points that lead to errors and decrease the efficiency of the classification.

\subsection{SIDER Dataset}
\begin{figure*}[t!]
\centering
\includegraphics[width=\linewidth]{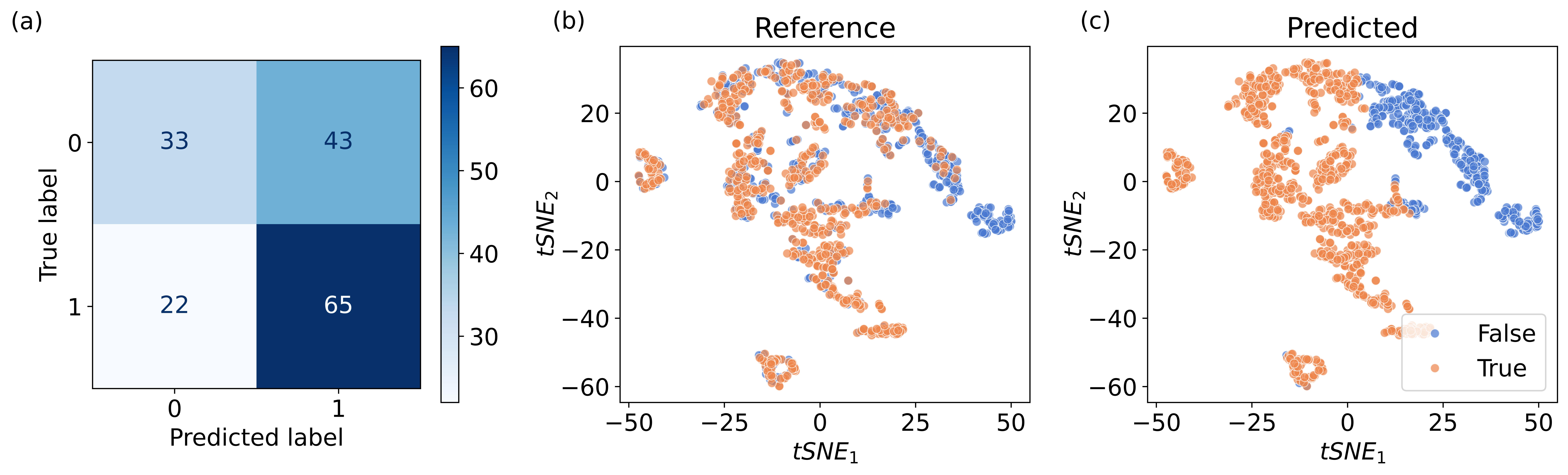}
\caption{\textbf{Classification of SIDER-1 task:} a) Confusion matrix of predicted labels on the test set. b) t-SNE feature representation of the entire reference dataset according to their labels. c) Feature representation of the predicted labels. }
\label{fig:sider} 
\end{figure*}

The SIDER dataset serves as a comprehensive pharmacovigilance resource, containing structured information on drug-associated side effects. Curated from diverse sources such as clinical trials, regulatory reports, and medical literature, it offers a systematic compilation of adverse drug reactions associated with various pharmaceutical interventions. The training results are presented in Table \ref{tab:sider}. Fig. \ref{fig:sider} illustrates label separation in the feature space for the SIDER-1 task. 
\begin{table}[h]
\caption{ROC-AUC performances of V0 and V1 models on the SIDER dataset. MMGNN denotes the prior top-performing results.}
\centering
\begin{tabular}{lccccccccc}
\hline
\hline
SIDER N & 1    & 2    & 3    & 4    & 5    & 6    & 7    & 8    & 9      \\
\hline
V0      & 0.684 & 0.709 & 0.985 & 0.676 & 0.847 & 0.738 & 0.945 & 0.839 & 0.739   \\
V1      & 0.663 & 0.707 & 0.983 & 0.689 & 0.838 & 0.744 & 0.941 & 0.832 & 0.724 \\
MMGNN   & 0.754 & 0.693 & 0.723 & 0.744 & 0.817 & 0.741 & -    & -    & -       \\
\hline
SIDER N & 10   & 11   & 12   & 13   & 14  & 15   & 16   & 17   & 18     \\
\hline
V0     & 0.775 & 0.908 & 0.819 & 0.853 & 0.809 & 0.723 & 0.926 & 0.822 & 0.724  \\
V1    &  0.760 & 0.892 & 0.816 & 0.849  & 0.782 & 0.737 & 0.917 & 0.833 & 0.736  \\
MMGNN   & -    & -    & -    & -    & -    & -    & -    & -    & -   \\
\hline
SIDER N & 19   & 20   & 21   & 22   & 23   & 24   & 25   & 26   & 27 \\
\hline
V0      & 0.724 & 0.759 & 0.764 & 0.700 & 0.927 & 0.598 & 0.747 & 0.925 & 0.698 \\
V1      & 0.736 & 0.772 & 0.726 & 0.708 & 0.923 & 0.599 & 0.742 & 0.916 & 0.684  \\
MMGNN   & -    & -    & -    & -    & -    & -   & -  & -  &  -   \\
\hline
\end{tabular}
\label{tab:sider}
\end{table}

\section{Position encoding}
\label{sec:appendixC}
\begin{wraptable}[10]{r}{0.41\linewidth}
\vspace{-1.25 cm}
\caption{ROC-AUC values for the V1 model with (P$^+$) and without positional encoding (P$^-$).}
\begin{tabular}{ccc}
\hline
BENCHMARK       & P$^+$    & P$^-$ \\
\hline
MP METALLICITY  & 0.965 & 0.966 \\
SG              & 0.968 & 0.966 \\
LA              & 0.980 & 0.969 \\
DIM             & 0.958 & 0.958 \\
BACE            & 0.856 & 0.808 \\
BBBP            & 0.905 & 0.888 \\
ClinTox         & 0.951 & 0.946 \\
HIV             & 0.979 & 0.980 \\
\hline
\end{tabular}
\label{tab:benchmark_p}
\end{wraptable} 
The order of atoms in structures could sometimes be important, especially if a set of structures was generated or described in the same way. If the input sequence follows specific rules and order, positional encoding can capture additional information and enhance prediction quality. Comparing our property predictions with and without positional embeddings, we find that the former indeed enhances the accuracy of the predictions. The results are outlined in Table \ref{tab:benchmark_p}.

\end{appendices}
\clearpage
\bibliographystyle{unsrtnat}
\bibliography{references}

\begin{thebibliography}{50}
\providecommand{\natexlab}[1]{#1}
\providecommand{\url}[1]{\texttt{#1}}
\expandafter\ifx\csname urlstyle\endcsname\relax
  \providecommand{\doi}[1]{doi: #1}\else
  \providecommand{\doi}{doi: \begingroup \urlstyle{rm}\Url}\fi

\bibitem[Ng et~al.(2020)Ng, Zhao, Yan, Conduit, and Seh]{Ng2020Predicting}
Man-Fai Ng, Jin Zhao, Qingyu Yan, Gareth~J Conduit, and Zhi~Wei Seh.
\newblock Predicting the state of charge and health of batteries using
  data-driven machine learning.
\newblock \emph{Nature Machine Intelligence}, 2\penalty0 (3):\penalty0
  161--170, 2020.
\newblock ISSN 2522-5839.
\newblock \doi{10.1038/s42256-020-0156-7}.

\bibitem[Liu et~al.(2020)Liu, Guo, Zou, Li, and Shi]{Liu2020Machine}
Yue Liu, Biru Guo, Xinxin Zou, Yajie Li, and Siqi Shi.
\newblock Machine learning assisted materials design and discovery for
  rechargeable batteries.
\newblock \emph{Energy Storage Materials}, 31:\penalty0 434--450, 2020.
\newblock ISSN 2405-8297.
\newblock \doi{10.1016/j.ensm.2020.06.033}.

\bibitem[Sawant et~al.(2023)Sawant, Deshmukh, and Awati]{Sawant2023Machine}
Vaishali Sawant, Rashmi Deshmukh, and Chetan Awati.
\newblock Machine learning techniques for prediction of capacitance and
  remaining useful life of supercapacitors: A comprehensive review.
\newblock \emph{Journal of Energy Chemistry}, 77:\penalty0 438--451, 2023.
\newblock ISSN 2095-4956.
\newblock \doi{10.1016/j.jechem.2022.11.012}.

\bibitem[Iwasaki et~al.(2019)Iwasaki, Takeuchi, Stanev, Kusne, Ishida,
  Kirihara, Ihara, Sawada, Terashima, Someya, Uchida, Saitoh, and
  Yorozu]{Iwasaki2019Machine-learning}
Yuma Iwasaki, Ichiro Takeuchi, Valentin Stanev, Aaron~Gilad Kusne, Masahiko
  Ishida, Akihiro Kirihara, Kazuki Ihara, Ryohto Sawada, Koichi Terashima,
  Hiroko Someya, Ken~ichi Uchida, Eiji Saitoh, and Shinichi Yorozu.
\newblock Machine-learning guided discovery of a new thermoelectric material.
\newblock \emph{Scientific Reports}, 9\penalty0 (1):\penalty0 1--7, 2019.
\newblock ISSN 20452322.
\newblock \doi{10.1038/s41598-019-39278-z}.

\bibitem[Akhter et~al.(2019)Akhter, Mekhilef, Mokhlis, and
  Shah]{Akhter2019Review}
Muhammad~Naveed Akhter, Saad Mekhilef, Hazlie Mokhlis, and Noraisyah~Mohamed
  Shah.
\newblock Review on forecasting of photovoltaic power generation based on
  machine learning and metaheuristic techniques.
\newblock \emph{IET Renewable Power Generation}, 13\penalty0 (7):\penalty0
  1009--1023, 2019.
\newblock ISSN 17521424.
\newblock \doi{10.1049/iet-rpg.2018.5649}.

\bibitem[Toyao et~al.(2020)Toyao, Maeno, Takakusagi, Kamachi, Takigawa, and
  Shimizu]{Toyao2020Machine}
Takashi Toyao, Zen Maeno, Satoru Takakusagi, Takashi Kamachi, Ichigaku
  Takigawa, and Ken-ichi Shimizu.
\newblock Machine learning for catalysis informatics: Recent applications and
  prospects.
\newblock \emph{ACS Catalysis}, 10\penalty0 (3):\penalty0 2260--2297, 2 2020.
\newblock \doi{10.1021/acscatal.9b04186}.

\bibitem[Vamathevan et~al.(2019)Vamathevan, Clark, Czodrowski, Dunham, Ferran,
  Lee, Li, Madabhushi, Shah, Spitzer, and Zhao]{Vamathevan2019Applications}
Jessica Vamathevan, Dominic Clark, Paul Czodrowski, Ian Dunham, Edgardo Ferran,
  George Lee, Bin Li, Anant Madabhushi, Parantu Shah, Michaela Spitzer, and
  Shanrong Zhao.
\newblock Applications of machine learning in drug discovery and development.
\newblock \emph{Nature Reviews Drug Discovery}, 18\penalty0 (6):\penalty0
  463--477, 2019.
\newblock ISSN 14741784.
\newblock \doi{10.1038/s41573-019-0024-5}.

\bibitem[Mikolov et~al.(2013)Mikolov, Chen, Corrado, and
  Dean]{Mikolov2013Efficient}
Tomas Mikolov, Kai Chen, Greg Corrado, and Jeffrey Dean.
\newblock Efficient estimation of word representations in vector space.
\newblock 2013.
\newblock \doi{10.48550/arXiv.1301.3781}.

\bibitem[Tshitoyan et~al.(2019)Tshitoyan, Dagdelen, Weston, Dunn, Rong,
  Kononova, Persson, Ceder, and Jain]{Tshitoyan2019Unsupervised}
Vahe Tshitoyan, John Dagdelen, Leigh Weston, Alexander Dunn, Ziqin Rong, Olga
  Kononova, Kristin~A. Persson, Gerbrand Ceder, and Anubhav Jain.
\newblock Unsupervised word embeddings capture latent knowledge from materials
  science literature.
\newblock \emph{Nature}, 571\penalty0 (7763):\penalty0 95--98, 2019.
\newblock ISSN 14764687.
\newblock \doi{10.1038/s41586-019-1335-8}.

\bibitem[Hansen et~al.(2015)Hansen, Biegler, Ramakrishnan, Pronobis,
  Von~Lilienfeld, Müller, and Tkatchenko]{Hansen2015Machine}
Katja Hansen, Franziska Biegler, Raghunathan Ramakrishnan, Wiktor Pronobis,
  O.~Anatole Von~Lilienfeld, Klaus~Robert Müller, and Alexandre Tkatchenko.
\newblock Machine learning predictions of molecular properties: Accurate
  many-body potentials and nonlocality in chemical space.
\newblock \emph{Journal of Physical Chemistry Letters}, 6\penalty0
  (12):\penalty0 2326--2331, 2015.
\newblock ISSN 19487185.
\newblock \doi{10.1021/acs.jpclett.5b00831}.

\bibitem[Jaeger et~al.(2018)Jaeger, Fulle, and Turk]{Jaeger2018Mol2vec}
Sabrina Jaeger, Simone Fulle, and Samo Turk.
\newblock Mol2vec: Unsupervised machine learning approach with chemical
  intuition.
\newblock \emph{Journal of Chemical Information and Modeling}, 58\penalty0
  (1):\penalty0 27--35, 2018.
\newblock ISSN 15205142.
\newblock \doi{10.1021/acs.jcim.7b00616}.

\bibitem[Goh et~al.(2018)Goh, Hodas, Siegel, and Vishnu]{Goh2017SMILES2Vec}
Garrett~B. Goh, Nathan~O. Hodas, Charles Siegel, and Abhinav Vishnu.
\newblock Smiles2vec: An interpretable general-purpose deep neural network for
  predicting chemical properties.
\newblock 2018.
\newblock \doi{10.48550/arXiv.1712.02034}.

\bibitem[Zhang et~al.(2020{\natexlab{a}})Zhang, Wang, Kaushik, Chu, Shan, Zhao,
  Xu, and Wei]{Zhang2020SPVec}
Yu~Fang Zhang, Xiangeng Wang, Aman~Chandra Kaushik, Yanyi Chu, Xiaoqi Shan,
  Ming~Zhu Zhao, Qin Xu, and Dong~Qing Wei.
\newblock Spvec: A word2vec-inspired feature representation method for
  drug-target interaction prediction.
\newblock \emph{Frontiers in Chemistry}, 7\penalty0 (January):\penalty0 1--11,
  2020{\natexlab{a}}.
\newblock ISSN 22962646.
\newblock \doi{10.3389/fchem.2019.00895}.

\bibitem[Stanev et~al.(2018)Stanev, Oses, Kusne, Rodriguez, Paglione,
  Curtarolo, and Takeuchi]{Stanev2018Machine}
Valentin Stanev, Corey Oses, A.~Gilad Kusne, Efrain Rodriguez, Johnpierre
  Paglione, Stefano Curtarolo, and Ichiro Takeuchi.
\newblock Machine learning modeling of superconducting critical temperature.
\newblock \emph{npj Computational Materials}, 4\penalty0 (1), 2018.
\newblock ISSN 20573960.
\newblock \doi{10.1038/s41524-018-0085-8}.

\bibitem[Chen et~al.(2019)Chen, Ye, Zuo, Zheng, and Ong]{Chen2019Graph}
Chi Chen, Weike Ye, Yunxing Zuo, Chen Zheng, and Shyue~Ping Ong.
\newblock Graph networks as a universal machine learning framework for
  molecules and crystals.
\newblock \emph{Chemistry of Materials}, 31\penalty0 (9):\penalty0 3564--3572,
  5 2019.
\newblock ISSN 15205002.
\newblock \doi{10.1021/acs.chemmater.9b01294}.

\bibitem[Kong et~al.(2022)Kong, Ricci, Guevarra, Neaton, Gomes, and
  Gregoire]{Kong2022Density}
Shufeng Kong, Francesco Ricci, Dan Guevarra, Jeffrey~B. Neaton, Carla~P. Gomes,
  and John~M. Gregoire.
\newblock Density of states prediction for materials discovery via contrastive
  learning from probabilistic embeddings.
\newblock \emph{Nature Communications}, 13\penalty0 (1):\penalty0 1--12, 2022.
\newblock ISSN 20411723.
\newblock \doi{10.1038/s41467-022-28543-x}.

\bibitem[Gori et~al.(2005)Gori, Monfardini, and Scarselli]{Gori2005new}
Marco Gori, Gabriele Monfardini, and Franco Scarselli.
\newblock A new model for earning in raph domains.
\newblock \emph{Proceedings of the International Joint Conference on Neural
  Networks}, 2:\penalty0 729--734, 2005.
\newblock \doi{10.1109/IJCNN.2005.1555942}.

\bibitem[Zhang et~al.(2020{\natexlab{b}})Zhang, Liu, and
  Xie]{Zhang2020Molecular}
Shuo Zhang, Yang Liu, and Lei Xie.
\newblock Molecular mechanics-driven graph neural network with multiplex graph
  for molecular structures.
\newblock \emph{arXiv}, pages 1--14, 2020{\natexlab{b}}.
\newblock \doi{10.48550/arXiv.2011.07457}.

\bibitem[Shui and Karypis(2020)]{Shui2020Heterogeneous}
Zeren Shui and George Karypis.
\newblock Heterogeneous molecular graph neural networks for predicting molecule
  properties.
\newblock \emph{Proceedings - IEEE International Conference on Data Mining,
  ICDM}, 2020-Novem:\penalty0 492--500, 2020.
\newblock ISSN 15504786.
\newblock \doi{10.1109/ICDM50108.2020.00058}.

\bibitem[Fung et~al.(2021)Fung, Zhang, Juarez, and
  Sumpter]{Fung2021Benchmarking}
Victor Fung, Jiaxin Zhang, Eric Juarez, and Bobby~G. Sumpter.
\newblock Benchmarking graph neural networks for materials chemistry.
\newblock \emph{npj Computational Materials}, 7\penalty0 (1):\penalty0 1--8,
  2021.
\newblock ISSN 20573960.
\newblock \doi{10.1038/s41524-021-00554-0}.

\bibitem[Xie and Grossman(2018)]{Xie2018Crystal}
Tian Xie and Jeffrey~C. Grossman.
\newblock Crystal graph convolutional neural networks for an accurate and
  interpretable prediction of material properties.
\newblock \emph{Physical Review Letters}, 120\penalty0 (14):\penalty0 145301, 4
  2018.
\newblock ISSN 10797114.
\newblock \doi{10.1103/PhysRevLett.120.145301}.

\bibitem[Guo et~al.(2021)Guo, Zhang, Yu, Herr, Wiest, Jiang, and
  Chawla]{Guo2021Few-shot}
Zhichun Guo, Chuxu Zhang, Wenhao Yu, John Herr, Olaf Wiest, Meng Jiang, and
  V.~Nitesh Chawla.
\newblock Few-shot graph learning for molecular property prediction.
\newblock \emph{The Web Conference 2021 - Proceedings of the World Wide Web
  Conference, WWW 2021}, pages 2559--2567, 2021.
\newblock \doi{10.1145/3442381.3450112}.

\bibitem[Bartók et~al.(2013)Bartók, Kondor, and Csányi]{Bartók2013On}
Albert~P. Bartók, Risi Kondor, and Gábor Csányi.
\newblock On representing chemical environments.
\newblock \emph{Physical Review B - Condensed Matter and Materials Physics},
  87\penalty0 (18):\penalty0 1--16, 2013.
\newblock ISSN 10980121.
\newblock \doi{10.1103/PhysRevB.87.184115}.

\bibitem[Huo and Rupp(2022)]{Huo2022Unified}
Haoyan Huo and Matthias Rupp.
\newblock Unified representation of molecules and crystals for machine
  learning.
\newblock \emph{Machine Learning: Science and Technology}, 3:\penalty0 045017,
  2022.
\newblock ISSN 26322153.
\newblock \doi{10.1088/2632-2153/aca005}.

\bibitem[Behler(2021)]{Behler2021Four}
Jörg Behler.
\newblock Four generations of high-dimensional neural network potentials.
\newblock \emph{Chemical Reviews}, 121:\penalty0 10037--10072, 2021.
\newblock ISSN 15206890.
\newblock \doi{10.1021/acs.chemrev.0c00868}.

\bibitem[Unke and Meuwly(2019)]{Unke2019PhysNet}
Oliver~T. Unke and Markus Meuwly.
\newblock Physnet: A neural network for predicting energies, forces, dipole
  moments, and partial charges.
\newblock \emph{Journal of Chemical Theory and Computation}, 15\penalty0
  (6):\penalty0 3678--3693, 2019.
\newblock ISSN 15499626.
\newblock \doi{10.1021/acs.jctc.9b00181}.

\bibitem[Vaswani et~al.(2023)Vaswani, Shazeer, Parmar, Uszkoreit, Jones, Gomez,
  Kaiser, and Polosukhin]{Vaswani2017Attention}
Ashish Vaswani, Noam Shazeer, Niki Parmar, Jakob Uszkoreit, Llion Jones,
  Aidan~N. Gomez, Lukasz Kaiser, and Illia Polosukhin.
\newblock Attention is all you need.
\newblock 2023.
\newblock \doi{10.48550/arXiv.1712.02034}.

\bibitem[Louis et~al.(2020)Louis, Zhao, Nasiri, Wang, Song, Liu, and
  Hu]{Louis2020Graph}
Steph~Yves Louis, Yong Zhao, Alireza Nasiri, Xiran Wang, Yuqi Song, Fei Liu,
  and Jianjun Hu.
\newblock Graph convolutional neural networks with global attention for
  improved materials property prediction.
\newblock \emph{Physical Chemistry Chemical Physics}, 22\penalty0
  (32):\penalty0 18141--18148, 2020.
\newblock ISSN 14639076.
\newblock \doi{10.1039/d0cp01474e}.

\bibitem[Billinge(2019)]{Billinge2019rise}
Simon~J.L. Billinge.
\newblock The rise of the x-ray atomic pair distribution function method: A
  series of fortunate events.
\newblock \emph{Philosophical Transactions of the Royal Society A:
  Mathematical, Physical and Engineering Sciences}, 377\penalty0 (2147), 2019.
\newblock ISSN 1364503X.
\newblock \doi{10.1098/rsta.018.0413}.

\bibitem[Hjorth~Larsen et~al.(2017)Hjorth~Larsen, Jørgen~Mortensen, Blomqvist,
  Castelli, Christensen, Dułak, Friis, Groves, Hammer, Hargus, Hermes,
  Jennings, Bjerre~Jensen, Kermode, Kitchin, Leonhard~Kolsbjerg, Kubal,
  Kaasbjerg, Lysgaard, Bergmann~Maronsson, Maxson, Olsen, Pastewka, Peterson,
  Rostgaard, Schiøtz, Schütt, Strange, Thygesen, Vegge, Vilhelmsen, Walter,
  Zeng, and Jacobsen]{Larsen2017atomic}
Ask Hjorth~Larsen, Jens Jørgen~Mortensen, Jakob Blomqvist, Ivano~E Castelli,
  Rune Christensen, Marcin Dułak, Jesper Friis, Michael~N Groves, Bjørk
  Hammer, Cory Hargus, Eric~D Hermes, Paul~C Jennings, Peter Bjerre~Jensen,
  James Kermode, John~R Kitchin, Esben Leonhard~Kolsbjerg, Joseph Kubal,
  Kristen Kaasbjerg, Steen Lysgaard, Jón Bergmann~Maronsson, Tristan Maxson,
  Thomas Olsen, Lars Pastewka, Andrew Peterson, Carsten Rostgaard, Jakob
  Schiøtz, Ole Schütt, Mikkel Strange, Kristian~S Thygesen, Tejs Vegge, Lasse
  Vilhelmsen, Michael Walter, Zhenhua Zeng, and Karsten~W Jacobsen.
\newblock The atomic simulation environment—a python library for working with
  atoms.
\newblock \emph{Journal of Physics: Condensed Matter}, 29\penalty0
  (27):\penalty0 273002, 2017.
\newblock ISSN 0953-8984.
\newblock \doi{10.1088/1361-648X/aa680e}.

\bibitem[Shermukhamedov et~al.(2024)Shermukhamedov, Mamurjonova, and
  Probst]{Shermukhamedov2023Structure}
Shokirbek Shermukhamedov, Dilorom Mamurjonova, and Michael Probst.
\newblock Structure to property: Chemical element embeddings and a deep
  learning approach for accurate prediction of chemical properties.
\newblock \emph{J. Chem. Inf. Model.}, 64\penalty0 (16):\penalty0 5762–5770,
  2024.
\newblock \doi{10.1021/acs.jcim.3c01990}.

\bibitem[Damm et~al.(1997)Damm, Frontera, Rives, and Jorgensen]{Damm1997OPLS}
Wolfgang Damm, Antonio Frontera, Julian~Tirado Rives, and William~L Jorgensen.
\newblock Opls all-atom force field for carbohydrates.
\newblock \emph{Journal of computational chemistry}, 18\penalty0 (16):\penalty0
  1955--1970, 1997.
\newblock \doi{10.1002/(SICI)1096-987X(199712)18:16<1955::AID-JCC1>3.0.CO;2-L}.

\bibitem[Gražulis et~al.(2009)Gražulis, Chateigner, Downs, Yokochi, Quirós,
  Lutterotti, Manakova, Butkus, Moeck, and
  Le~Bail]{Gražulis2009Crystallography}
Saulius Gražulis, Daniel Chateigner, Robert~T Downs, A~F~T Yokochi, Miguel
  Quirós, Luca Lutterotti, Elena Manakova, Justas Butkus, Peter Moeck, and
  Armel Le~Bail.
\newblock Crystallography open database – an open-access collection of
  crystal structures.
\newblock \emph{Journal of Applied Crystallography}, 42\penalty0 (4):\penalty0
  726--729, 8 2009.
\newblock \doi{10.1107/S0021889809016690}.

\bibitem[Gražulis et~al.(2012)Gražulis, Daškevič, Merkys, Chateigner,
  Lutterotti, Quirós, Serebryanaya, Moeck, Downs, and
  Le~Bail]{Gražulis2012Crystallography}
Saulius Gražulis, Adriana Daškevič, Andrius Merkys, Daniel Chateigner, Luca
  Lutterotti, Miguel Quirós, Nadezhda~R Serebryanaya, Peter Moeck, Robert~T
  Downs, and Armel Le~Bail.
\newblock Crystallography open database (cod): an open-access collection of
  crystal structures and platform for world-wide collaboration.
\newblock \emph{Nucleic Acids Research}, 40\penalty0 (D1):\penalty0 D420--D427,
  1 2012.
\newblock ISSN 0305-1048.
\newblock \doi{10.1093/nar/gkr900}.

\bibitem[Jain et~al.(2013)Jain, Ong, Hautier, Chen, Richards, Dacek, Cholia,
  Gunter, Skinner, Ceder, and Persson]{Jain2013Commentary}
Anubhav Jain, Shyue~Ping Ong, Geoffroy Hautier, Wei Chen, William~Davidson
  Richards, Stephen Dacek, Shreyas Cholia, Dan Gunter, David Skinner, Gerbrand
  Ceder, and Kristin~A. Persson.
\newblock Commentary: The materials project: A materials genome approach to
  accelerating materials innovation.
\newblock \emph{APL Materials}, 1\penalty0 (1), 2013.
\newblock ISSN 2166532X.
\newblock \doi{10.1063/1.4812323}.

\bibitem[Ong et~al.(2015)Ong, Cholia, Jain, Brafman, Gunter, Ceder, and
  Persson]{Ong2015Materials}
Shyue~Ping Ong, Shreyas Cholia, Anubhav Jain, Miriam Brafman, Dan Gunter,
  Gerbrand Ceder, and Kristin~A. Persson.
\newblock The materials application programming interface (api): A simple,
  flexible and efficient api for materials data based on representational state
  transfer (rest) principles.
\newblock \emph{Computational Materials Science}, 97:\penalty0 209--215, 2015.
\newblock ISSN 09270256.
\newblock \doi{10.1016/j.commatsci.2014.10.037}.

\bibitem[Banik et~al.(2023)Banik, Dhabal, Chan, Manna, Cherukara, Molinero, and
  Sankaranarayanan]{Banik2023CEGANN}
Suvo Banik, Debdas Dhabal, Henry Chan, Sukriti Manna, Mathew Cherukara, Valeria
  Molinero, and Subramanian~K.R.S. Sankaranarayanan.
\newblock Cegann: Crystal edge graph attention neural network for multiscale
  classification of materials environment.
\newblock \emph{npj Computational Materials}, 9\penalty0 (1):\penalty0 1--12,
  2023.
\newblock ISSN 20573960.
\newblock \doi{10.1038/s41524-023-00975-z}.

\bibitem[Ziletti et~al.(2018)Ziletti, Kumar, Scheffler, and
  Ghiringhelli]{Ziletti2018Insightful}
Angelo Ziletti, Devinder Kumar, Matthias Scheffler, and Luca~M. Ghiringhelli.
\newblock Insightful classification of crystal structures using deep learning.
\newblock \emph{Nature Communications}, 9\penalty0 (1):\penalty0 1--10, 2018.
\newblock ISSN 20411723.
\newblock \doi{10.1038/s41467-018-05169-6}.

\bibitem[Wu et~al.(2018)Wu, Ramsundar, Feinberg, Gomes, Geniesse, Pappu,
  Leswing, and Pande]{Wu2018MoleculeNet}
Zhenqin Wu, Bharath Ramsundar, Evan~N. Feinberg, Joseph Gomes, Caleb Geniesse,
  Aneesh~S. Pappu, Karl Leswing, and Vijay Pande.
\newblock Moleculenet: A benchmark for molecular machine learning.
\newblock \emph{Chemical Science}, 9\penalty0 (2):\penalty0 513--530, 2018.
\newblock ISSN 20416539.
\newblock \doi{10.1039/c7sc02664a}.

\bibitem[Kuhn et~al.(2016)Kuhn, Letunic, Jensen, and Bork]{Kuhn2016SIDER}
Michael Kuhn, Ivica Letunic, Lars~Juhl Jensen, and Peer Bork.
\newblock The sider database of drugs and side effects.
\newblock \emph{Nucleic Acids Research}, 44\penalty0 (D1):\penalty0
  D1075--D1079, 2016.
\newblock ISSN 13624962.
\newblock \doi{10.1093/nar/gkv1075}.

\bibitem[O'Boyle et~al.(2011)O'Boyle, Banck, James, Morley, Vandermeersch, and
  Hutchison]{O'Boyle2011Open}
Noel~M O'Boyle, Michael Banck, Craig~A James, Chris Morley, Tim Vandermeersch,
  and Geoffrey~R Hutchison.
\newblock Open babel: An open chemical toolbox.
\newblock \emph{Journal of Cheminformatics}, 3\penalty0 (33):\penalty0 1--14,
  2011.
\newblock ISSN 17582946.
\newblock \doi{10.1186/1758-2946-3-33}.

\bibitem[RDK()]{RDKit}
Rdkit: Open-source cheminformatics.
\newblock URL \url{https://www.rdkit.org}.

\bibitem[Chollet et~al.(2015)]{Chollet2015Keras}
Fran\c{c}ois Chollet et~al.
\newblock Keras.
\newblock 2015.
\newblock URL \url{https://keras.io}.

\bibitem[Chen and Ong(2021)]{Chen2021AtomSets}
Chi Chen and Shyue~Ping Ong.
\newblock Atomsets as a hierarchical transfer learning framework for small and
  large materials datasets.
\newblock \emph{npj Computational Materials}, 7\penalty0 (1), 2021.
\newblock ISSN 20573960.
\newblock \doi{10.1038/s41524-021-00639-w}.

\bibitem[Li et~al.(2022)Li, Hsieh, Lu, Gong, Wang, Li, Liu, Tian, Jiang, Yan,
  Bai, Liu, Zhang, and Yao]{Li2022GLAM}
Yuquan Li, Chang-Yu Hsieh, Ruiqiang Lu, Xiaoqing Gong, Xiaorui Wang, Pengyong
  Li, Shuo Liu, Yanan Tian, Dejun Jiang, Jiaxian Yan, Qifeng Bai, Huanxiang
  Liu, Shengyu Zhang, and Xiaojun Yao.
\newblock Glam : An adaptive graph learning method for automated molecular
  interactions and properties predictions.
\newblock \emph{Nature Machine Intelligence}, 4\penalty0 (7):\penalty0
  645--651, 2022.
\newblock ISSN 2522-5839.
\newblock \doi{10.1038/s42256-022-00501-8}.

\bibitem[Li et~al.(2021)Li, Li, Hsieh, Zhang, Liu, Liu, Song, and
  Yao]{Li2021TrimNet}
Pengyong Li, Yuquan Li, Chang-Yu Hsieh, Shengyu Zhang, Xianggen Liu, Huanxiang
  Liu, Sen Song, and Xiaojun Yao.
\newblock Trimnet: learning molecular representation from triplet messages for
  biomedicine.
\newblock \emph{Briefings in Bioinformatics}, 22\penalty0 (4):\penalty0
  bbaa266, 7 2021.
\newblock ISSN 1477-4054.
\newblock \doi{10.1093/bib/bbaa266}.

\bibitem[Baek et~al.(2021)Baek, Kang, and Hwang]{Baek2021Accurate}
Jinheon Baek, Minki Kang, and Sung~Ju Hwang.
\newblock Accurate learning of graph representations with graph multiset
  pooling.
\newblock 2021.
\newblock \doi{10.48550/arXiv.2102.11533}.

\bibitem[Fabian et~al.(2020)Fabian, Edlich, Gaspar, Segler, Meyers, Fiscato,
  and Ahmed]{Fabian2020Molecular}
Benedek Fabian, Thomas Edlich, Héléna Gaspar, Marwin Segler, Joshua Meyers,
  Marco Fiscato, and Mohamed Ahmed.
\newblock Molecular representation learning with language models and
  domain-relevant auxiliary tasks.
\newblock 2020.
\newblock \doi{10.48550/arXiv.2011.13230}.

\bibitem[Irwin et~al.(2022)Irwin, Dimitriadis, He, and
  Bjerrum]{Irwin2022Chemformer}
Ross Irwin, Spyridon Dimitriadis, Jiazhen He, and Esben~Jannik Bjerrum.
\newblock Chemformer: A pre-trained transformer for computational chemistry.
\newblock \emph{Machine Learning: Science and Technology}, 3\penalty0 (1),
  2022.
\newblock ISSN 26322153.
\newblock \doi{10.1088/2632-2153/ac3ffb}.

\bibitem[Pedregosa et~al.(2011)Pedregosa, Varoquaux, Gramfort, Michel, Thirion,
  Grisel, Blondel, Prettenhofer, Weiss, Dubourg, Vanderplas, Passos,
  Cournapeau, Brucher, Perrot, and {{\'E}}douard
  Duchesnay]{Pedregosa2011Scikit-learn}
Fabian Pedregosa, Ga{{\"e}}l Varoquaux, Alexandre Gramfort, Vincent Michel,
  Bertrand Thirion, Olivier Grisel, Mathieu Blondel, Peter Prettenhofer, Ron
  Weiss, Vincent Dubourg, Jake Vanderplas, Alexandre Passos, David Cournapeau,
  Matthieu Brucher, Matthieu Perrot, and {{\'E}}douard Duchesnay.
\newblock Scikit-learn: Machine learning in python.
\newblock \emph{Journal of Machine Learning Research}, 12\penalty0
  (85):\penalty0 2825--2830, 2011.
\newblock URL \url{http://jmlr.org/papers/v12/pedregosa11a.html}.

\end{thebibliography}

\end{document}